# Mapping complex cell morphology in the grey matter with double diffusion encoding MR: a simulation study

A. Ianuş[1,2], D.C. Alexander[1], H. Zhang[1], M. Palombo[1*]

[1]Centre for Medical Image Computing and Department of Computer Science, University College London, London, UK

[2]Champalimaud Research, Champalimaud Centre for the Unknown, Lisbon, Portugal

*Corresponding author: Marco Palombo, e-mail address: marco.palombo@ucl.ac.uk.

## Abstract

This paper investigates the impact of cell body (namely soma) size and branching of cellular projections on diffusion MR imaging (dMRI) and spectroscopy (dMRS) signals for both standard single diffusion encoding (SDE) and more advanced double diffusion encoding (DDE) measurements using numerical simulations. The aim is to investigate the ability of dMRI/dMRS to characterize the complex morphology of brain cells focusing on these two distinctive features of brain grey matter.

To this end, we employ a recently developed computational framework to create three dimensional meshes of neuron-like structures for Monte Carlo simulations, using diffusion coefficients typical of water and brain metabolites. Modelling the cellular structure as realistically connected spherical soma and cylindrical cellular projections, we cover a wide range of combinations of sphere radii and branching order of cellular projections, characteristic of various grey matter cells. We assess the impact of spherical soma size and branching order on the b-value dependence of the SDE signal as well as the time dependence of the mean diffusivity (MD) and mean kurtosis (MK). Moreover, we also assess the impact of spherical soma size and branching order on the angular modulation of DDE signal at different mixing times, together with the mixing time dependence of the apparent microscopic anisotropy (μA), a promising contrast derived from DDE measurements.

The SDE results show that spherical soma size has a measurable impact on both the b-value dependence of the SDE signal and the MD and MK diffusion time dependence for both water and metabolites. On the other hand, we show that branching order has little impact on either, especially for water. In contrast, the DDE results show that spherical soma size has a measurable impact on the DDE signal's angular modulation at short mixing times and the branching order of cellular projections significantly impacts the mixing time dependence of the DDE signal's angular modulation as well as of the derived μA, for both water and metabolites.

Our results confirm that SDE based techniques may be sensitive to spherical soma size, and most importantly, show for the first time that DDE measurements may be more sensitive to the dendritic tree complexity (as parametrized by the branching order of cellular projections), paving the way for new ways of characterizing grey matter morphology, non-invasively using dMRS and potentially dMRI.

# 1. Introduction

Non-invasive mapping of brain cells morphology is a major focus in biomedical imaging research, as it can play a crucial role in the assessment of neurologic and psychiatric diseases which alter the tissue structure [1], for studying brain development [2], plasticity [3] or ageing [1]. Soma size and the tree configuration of cellular projections of neurons and glia are largely plastic properties which are directly affected in various pathologies. For instance, a decrease in neuronal soma size has been reported in subjects with bipolar disorder [4], while an increase in motoneuron soma size is present in amyotrophic lateral sclerosis [5]. Abnormalities and changes in the dendritic tree characterize a wide range of disorders [6], including a progressive loss of dendrites and spines in normal aging [7]. Changes in glial cells, such as astrocyte hypertrophy/atrophy characterized by an overall increase/decrease in cell size, also accompany various pathologies, from traumatic brain injury [8] to Alzheimer's disease [9]. This information is usually obtained based on histological imaging of tissue samples, which is highly invasive and can be performed only ex-vivo.

Significant efforts are made to estimate microscopic tissue features in-vivo using non-invasive imaging techniques, and especially diffusion Magnetic Resonance Imaging (dMRI), which uses magnetic field gradients to sensitise the measured signal to the displacement of probe molecules (usually water) in the tissue. Then, by modelling the relationship between neuronal configurations and the measured signal, microscopic tissue properties could be inferred from the dMRI measurements. Towards this goal, various signal representations and biophysical models have been developed to capture different features of the complex brain tissue [10-12].

The majority of techniques aimed at mapping brain microstructure properties, either based on signal representations [13-15] or biophysical models [10, 11], have been focused on white matter. In terms of biophysical modelling approaches, they usually describe simple geometries [12], with a focus on estimating tissue features such as intra-neurite volume fraction [16-19], axon diameter [20-22], neurite dispersion [17, 23, 24], and membrane permeability [25, 26]. Some techniques have also been applied for mapping microstructure in grey matter, mainly focusing on neurite dispersion [17, 23, 27], and more recent studies have shown the potential of mapping soma apparent density and size [28] as well as branching complexity using diffusion of metabolites [29-32] [33] [34, 35]. Most of these techniques use a collection of standard single diffusion encoding (SDE) measurements and simple geometric representations to describe the brain tissue, for instance cylinders to mimic axons or spheres to represent the soma. Although such models can provide an insight into the gross effects of different tissue features on the dMRI signal and are useful for optimising the acquisition parameters, they are an oversimplification of real configurations and limit the ability of extracting details about the tissue structure. Moreover, there is increasing evidence that going beyond the standard acquisition and employing multi-dimensional diffusion sequences, such as double diffusion encoding (DDE) [36-42], double oscillating diffusion encoding (DODE) [43-45] or q-space trajectory encoding (QTI) [27, 46, 47], can provide additional information about the tissue microstructure compared to SDE acquisitions.

Developing a meaningful biophysical model of diffusion in grey matter is very challenging [48]. In contrast to white matter, grey matter is comprised of packed cell bodies and cellular projections, such as neuronal dendrites and glial projections, that branch and densely weave together in random configurations. Furthermore, each branch can present undulations, curvature and secondary structures such as spines that add another layer of complexity to the biophysical modelling. Therefore, to address the challenge of modelling grey matter microstructure and/or to understand the contrast of various signal representations, it is crucial to know which features of the cellular structure have a measurable impact on the diffusion-weighted MR signal.

A powerful tool that can be used to find an answer to this still open question is computational modelling. Numerical phantoms, and especially those based on Monte Carlo (MC) simulations, e.g. [35, 49-53], allow control and flexibility, both in terms of the underlying diffusion substrates, as well as the acquisition sequences, offering the unique opportunity to perform *in silico* experiments targeting specific features of the tissue microstructure. So far, such techniques have been employed with a wide range of synthetic tissues, from simple substrates of parallel cylinders [50, 52], substrates including fibre dispersion [54-56] and multiple fibre populations [57], to realistic meshes based on electron microscopy images of real tissue [58-60]. Although MC based simulations of dMRI signal have been around for decades [61-63], they have been either rather simplistic and very flexible (e.g. parallel cylinders) or highly realistic, but very rigid (e.g. meshes based on microscopy). Only recently, computational frameworks for designing realistic neuronal meshes for MC dMRI simulations have been proposed [49, 53, 56]. Specifically, the framework presented in Palombo et al. [49] offers flexibility and control by employing a generative model to create realistic meshes which closely resemble a wide range of brain cells, from glia to neurons. The dMRI simulations presented in [49] show the signal differences between several cell types, however they do not study the specific effects of different microstructural properties on the signal.

The aim of this study is to systematically investigate the effect of subtle morphological features such as cell soma size (modelled as sphere) and branching order of cellular projections

(modelled as connected cylinders) on the diffusion properties measured with water dMRI and/or metabolite diffusion-weighted MR spectroscopy (dMRS). First, we study the effect of branching order and spherical soma size on the b-value and diffusion time dependence of the signal measured with standard SDE sequences. Then, we investigate the signature of branching order and spherical soma size on the signal measured with DDE sequences and we assess whether we can use DDE measurements to inform on cell complexity.

## 2. Methods

In this section we first describe the overall design of the simulation experiments and the details of the implementation, including the computational models of brain cells used together with the details of the MC simulation. Then, we explain the two sets of simulation experiments we performed to investigate the impact of branching and spherical soma size on both SDE and DDE measurements, using two different diffusivities to mimic intracellular water and metabolites diffusion, respectively. In this way, our results can be used to inform both water dMRI and metabolites dMRS experiments.

### 2.1. General simulation design

The aim of this work is to systematically investigate the effect of spherical soma size and branching order of cylindrical projections on the diffusion properties measured with dMRI/dMRS. Towards this goal, we use the generative model introduced in [49] to create synthetic neuron-like cell structures with controllable complex features. Specifically, the generative model allows the design of realistic virtual cell structures by defining twelve morphological features including the number of cell projections, $N_{proj}$, number of consecutive bifurcations, $N_b$, cell branch length, $L_b$, and diameter, $D_b$ (Figure 1a), soma realistically connected to the projections with controllable diameter, $D_s$ (Figure 1b), cell branch undulations and curvature (Figure 1c), as well as complex secondary features such as dendritic spines with controllable size and density (Figure 1d). Given our specific goal, we focus our analysis on synthetic cell structures like that reported in Figure 1b, where features other than branching and spherical soma, e.g. cell branch undulation/curvature and dendritic spines, are not incorporated by design. The effect of branching and spherical soma size are investigated by systematically varying $N_b$ and $D_s$, for two values of cell domain sizes L, mimicking small and big neural cells, and two values of intrinsic diffusivity D, mimicking water and metabolite diffusion.

However, isolating the relative contribution of branching and soma size from other confounding factors is still challenging. In particular, the signal fraction of restricted diffusion within the spherical soma can have a significant influence on the time-dependence of the measured signal, hence the virtual cell models must be designed to have the same volume fraction occupied by the spherical soma $v_s$. To keep $v_s$ constant when changing branching complexity (by changing $N_b$) and/or spherical soma size (by changing $D_s$), there are three basic strategies to choose from: i) adjusting $N_{proj}$; ii) adjusting the overall size of the cell domain L (which leads to changes in $L_b$), or iii) adjusting the diameter $D_b$ of cylindrical branches. Strategies i) and ii) would lead to cell structures rather unrealistic/unusual to be considered. For example, if $v_s = 30\%$ and $D_b = 0.5$ µm, when high branching (e.g. $N_b = 6$) and small soma (e.g. $D_s = 8$ µm) are chosen, the overall cell domain and $N_{proj}$ would need to be very small to keep the soma volume fraction constant (e.g. leading to $L_b = 12$ µm for $N_{proj} = 3$). In the other extreme, when $D_s = 20$ µm and $N_b = 1$, the overall cell domain and $N_{proj}$ would need to be very large (e.g. leading to $L_b = 490$ µm for $N_{proj} = 100$). In contrast, strategy iii) avoids such unrealistic configurations. Therefore, we adopt strategy iii) when designing our computational models of neuron-like cells, and adjust $D_b$.

Nonetheless, using strategy iii) still leaves some confounding effects, due to cases where $D_b$ is large enough to have a measurable impact as well as due to different probabilities of exchange between soma and branches as $D_b$ is varied. Nevertheless, the signature of $D_b$ on the measured dMRI signal is predictable and would impact only a few cases (those with low $N_b$ and large $D_s$), while the effect of exchange can be investigated by comparison with a compartment model which exhibits no exchange. For these reasons, when designing our computational models of brain cells we take care of keeping the branch diameter below 3 µm, where it has a minimal influence on the diffusion weighted signal, especially at moderate gradient strength and medium and long diffusion times [64, 65], which are more typical conditions in real pre-clinical and clinical settings.

*2.2. Implementation of computational models of synthetic neurons*

Specifically, we choose four target soma diameters $D_s = \{8, 12, 16, 20\}$ µm to cover a wide range of spherical soma sizes seen in various neuron types [66, 67] and four branching orders $N_b = \{1, 2, 4, 6\}$. When constructing the synthetic cell, each cylindrical projection bifurcates $N_b-1$ times, with $N_b = 1$ corresponding to non-branching projections, and the complexity of the synthetic cellular structure increases exponentially with $N_b$. Figure 1e) shows an example set of the configurations investigated. Furthermore, we choose a fixed cellular volume fraction occupied by the soma $v_s = 30\%$, as a typical value for most of the brain cell types. Each cell has $N_{proj} = 10$ cylindrical projections leaving the soma, which is the average number of projections encountered in several cell types, for example in pyramidal cells, motoneurons, stellate and chandelier neurons [49]. The first branch of each projection radiating from the soma is isotropically distributed in space. At each bifurcation, the subsequent directions of the two new segments are drawn randomly in space, with a 60º angle between segments. This value is the average bifurcation angle for most of the brain cells [49]. We choose two target cell domain sizes L = {400, 1000} µm corresponding to the typical sizes of most of glial cells (i.e. astrocytes, oligodendrocytes and majority of microglia) and small and large neurons domain, for example pyramidal cells and motoneuron, respectively [49]. Since the target cell domain radius is L/2 and equal to the average total process length, $L_b \times N_b$, we set the target $L_b$ as:

$$L_b = \frac{L}{2N_b} \quad [1]$$

Given the fixed soma volume fraction $\nu_s$, the target $D_b$ can be computed from the definition of $\nu_s$ as:

$$D_b = \sqrt{\frac{2}{3} \frac{\left(\frac{1}{\nu_s}-1\right) D_s^3}{L_b N_{proj} (2^{N_b}-1)}} \quad [2]$$

where we used the definition of $\nu_s$ as:

$$\nu_s = \frac{spherical\ soma\ vol}{sph\ soma\ vol + total\ cyl\ projections\ vol} = \frac{\frac{4}{3}\pi \left(\frac{D_s}{2}\right)^3}{\frac{4}{3}\pi \left(\frac{D_s}{2}\right)^3 + N_{proj}(2^{N_b}-1)\pi L_b \left(\frac{D_b}{2}\right)^2}$$

with $N_{proj}(2^{N_b} - 1)$ being the total number of cylindrical segments of length $L_b$ and diameter $D_b$ comprising each cellular projection.

Then, for each cell configuration the potential of water molecules to exchange between soma and branches is proportional to the ratio between the total cross-sectional area of the projections leaving the soma and the soma surface:

$$p_{ex} \sim \frac{N_{proj} D_b^2}{4 D_s^2} \quad [3]$$

Given the probabilistic nature of the synthetic cell generation, for each parameter combination we create 10 cell instances which are used to average the simulated diffusion signal. Previous investigations [34, 35, 49] have suggested that 10 cell instances are the minimum number to guarantee a standard deviation of the simulated normalized signal lower than 2%.

Three-dimensional surface meshes are then generated in Blender 2.79 using "metaballs" objects, as previously described in [49]. To make the MC simulations less computationally expensive, the resulting meshes are simplified and smoothed in Blender 2.79 using the "decimate" and "smooth" modifiers, leading to sparser surface meshes of ~$10^3$ triangular faces. Following this procedure, the effective sizes of soma ($D_s$) and branches ($D_b$) of the final mesh may slightly differ from the target ones. To precisely measure the effective values of $D_s$ and $D_b$, we consider 256 rays, each one along a different direction in space chosen by uniformly sampling a sphere centred in the middle of the soma or each branch. Then, we compute for each ray the ray-triangle intersection to determine which triangle of the cellular mesh the ray intersect. Subsequently, we compute the distance from the origin of the ray to the intersection point on that triangular face of the mesh. The minimum of the 256 distances computed in this way (one for each ray) is taken as effective radius of the corresponding spherical soma or cylindrical branch. The morphometric parameters, both the target ones as well as the effective values after the meshing procedure, are given in Table S1 in Supplementary Information.

*2.3 Monte Carlo simulations*

To simulate the diffusion signal, we employ the MC simulator in Camino [68] [50]. The source code has been slightly modified to allow the user to input the initial walker coordinates. This significantly reduces the time required for walker placement inside the cells with Camino's built-in algorithm which is designed for generic meshes and thus not optimised for the morphology of neuronal cells we model here.

For each mesh, the initial walker coordinates are carefully generated using a custom script in MATLAB (The Mathworks) to ensure that the number of walkers placed in each branch segment or soma is proportional to its volume fraction with respect to the whole cell. In the soma, the walkers are placed in a smaller concentric sphere with a diameter of $0.9 \cdot D_s$. To allow for the distribution of the spins to reach a steady state, the first 50 ms of the simulation are discarded. For each configuration, we verify that the number of spins in the soma indeed stabilise after this period, for both the investigated diffusivities.

The MC simulations are run with the following parameters: diffusivity D = {2, 0.5} μm²/ms corresponding to typical intracellular water and metabolites diffusivities, respectively; $10^4$ walkers for each cell instance, resulting in $10^5$ walkers for each configuration and a step duration of $\delta t = 0.1$ ms, according to the general guidelines in [50]. The step duration, together with D, determines the fixed step size $\delta r = (6 \cdot D \cdot \delta t)^{1/2}$. We choose this $\delta t$ as a trade-off between accuracy of the MC simulation and computational time. The chosen $\delta t$ is small enough to guarantee that the standard deviation of the simulated normalized signal over the 10 cell instances is lower than 2% for the fast diffusivity and less than 0.7% for the slow diffusivity.

*2.4 Simulated mechanisms of exchange*

Our simulations are focused on intracellular dynamics and corresponding intracellular dMRI/dMRS signal only. With our simulation design, we can access two possible mechanisms of exchange:

1. *soma-branch exchange*, that is the exchange of diffusing molecules between the spherical soma and the cylindrical projections of our synthetic cells;
2. *branch-branch exchange*, that is the exchange of diffusing molecules between one branch of a projection to another of our synthetic cells.

Exchange of diffusing molecules between intracellular and extracellular compartments, namely *intra-extracellular exchange*, is not considered in our simulations.

*2.5 Effect of soma size and branching order on SDE measurements*

In the first set of simulation experiments, we investigate the effect of spherical soma size and branching order on the diffusion signal of ideal SDE sequences. First, we investigate the impact on the signal b-value dependence, then the effect on the time dependence of the mean diffusivity (MD) and mean diffusional kurtosis (MK) indices derived from the second-order cumulant expansion of the signal.

*2.5.1 B-value dependence*

We investigate the effect of spherical soma size and branching order of cylindrical projections on the b-value dependence for ideal SDE sequences with short gradient duration $\delta_1 = 1$ ms, three different diffusion gradient pulse separation times $\Delta_1 = \{10, 30, 80\}$ ms and 28 b-values ranging from 0 to 60 ms/µm$^2$. For each parameter combination, we average the signal over 32 isotropically oriented gradient directions.

To better understand the b-value dependence and the impact of the exchange between branches and soma, we compare these signals with those from a theoretical compartment model which accounts for restricted diffusion inside spheres and isotropically oriented finite cylinders with the effective diameter given in Table S1, and a length equal to half the cell domain L. The restricted diffusion signal is computed according to the Gaussian Phase Distribution (GPD) approximation [69].

*2.5.2 MD and MK time dependence*

To study the impact of spherical soma size and branching order of cylindrical projections on the MD and MK time dependence, we consider ideal SDE sequences with a gradient pulse duration of $\delta_2 = 1$ ms, 3 different b-values, specifically $b_2 = \{0.5, 1, 2\}$ ms/µm² for simulations with high diffusivity D = 2 µm²/ms and $b_2 = \{2, 4, 8\}$ ms/µm² for small diffusivity D = 0.5 µm²/ms, and 35 diffusion gradient pulse separation times $\Delta_2$ per b value ranging from 1.1 to 2450 ms. As in section 2.4.1, the signal is averaged over 32 directions, then, MD and MK are computed by fitting the following equation:

$$\log(S) = -MD \cdot b + \frac{1}{6} MD^2 \cdot MK \cdot b^2, \quad [4]$$

where S is the normalized diffusion signal from the MC simulations. The estimated MD and MK values are also compared to the theoretical values based on the two-compartment model.

*2.6 Effect of soma size and branching order on DDE measurements*

In this second set of experiments, we study the effect of spherical soma size and branching order on DDE measurements, which have been suggested to provide additional contrast compared to SDE sequences, especially related to microscopic diffusion anisotropy (µA) [11, 70-72]. We hypothesise that DDE measurements would be more sensitive to the branching of cellular projections than SDE ones. Specifically, we hypothesise that the angular modulation of the DDE signal and the derived apparent µA index can quantify more directly the loss of correlation between subsequent diffusion directions due to spins diffusing from one branch to another, oriented in a different direction.

A typical DDE experiment encompasses two diffusion-weighting blocks separated by a mixing time $\tau_m$. The specific DDE design to map restricted diffusion (in the case of negligible/slow inter-compartmental exchange) is to study the angular modulation of the DDE signal as a function of the relative angle between the diffusion gradients of the two blocks at short mixing time $\tau_m$, with the difference between parallel and anti-parallel measurements reflecting the restriction size. On the other hand, the specific DDE design to map microscopic anisotropy is to study the angular modulation at long $\tau_m$ [11, 36, 37, 70-72], with the difference between parallel and orthogonal measurements reflecting microscopic anisotropy. In this experiment, compartments without shape anisotropy would lead to flat angular modulation of the DDE signal's amplitude at long $\tau_m$, while compartments with shape anisotropy would preserve a strong modulation [37, 38, 40, 44, 73, 74].

However, in neuron-like structures such as those considered here, different cellular compartments of different shapes, e.g. spherical soma and (branched) cylindrical projections, are interconnected and the MR probe molecules (either water or metabolites) can exchange between them during the interval $\tau_m$, losing correlation between subsequent diffusion directions. As a consequence, in case of cellular structures with branched projections, if there is a non-negligible fraction of diffusing molecules which move between one branch to another oriented in a different direction and/or between branches and the spherical soma, then we should measure a lower amplitude of the angular DDE signal modulation due to the resulting loss of correlation between subsequent diffusion directions, compared to the non-exchanging case.

Here we design simulation experiments to investigate whether and how the presence of spherical soma of different sizes and cylindrical projections with different branching orders impact the $\tau_m$ dependence of the DDE signal's angular modulation. We also study the effect on a rotationally invariant index of apparent µA [75], which nowadays is more commonly used in anisotropic tissue compared to the amplitude modulation of the signal.

*2.6.1 Mixing time dependence of the DDE signal's angular modulation*

With the first simulation experiment, we investigate the effect of mixing time $\tau_m$ on the amplitude of angular DDE signal modulation, where the relative angle between the two gradient pairs is varied between 0 and $2\pi$ radians. To this end, we keep the diffusion time of each block short, to ensure negligible exchange between spherical soma and branches and between different branches during the individual blocks, thus isolating the contribution of soma and dendritic tree complexity to the signal only during the mixing time. In this way, we expect to see what in SDE experiments is driven by diffusion time, being driven instead by mixing time in DDE. This allows us to explore a unique DDE feature (mixing time) while providing a fair comparison with SDE as a single timing parameter is varied. Specifically, we consider ideal DDE sequences with a gradient pulse duration $\delta_3 = 1$ ms, short diffusion time $\Delta_3 = 5$ ms, three b-values ($b_3 = \{1, 2, 4\}$ ms/µm² for D = 2 µm²/ms and $b_3 = \{4, 8, 16\}$ ms/µm² for D = 0.5 µm²/ms), $\tau_m$ varying between 1 and 200 ms, and the relative angle between the gradients φ is varied in 17 steps between 0 and $2\pi$ radians. These sequence parameters are chosen to ensure that the exchange between soma and branches and between different branches is negligible during the interval $\Delta_3$ (i.e. the root mean squared displacement along the branch $\leq$ 10% $L_{branch}$), but it can have a significant effect when increasing $\tau_m$.

To mitigate the effect of any residual macroscopic anisotropy, we use a scheme similar to the ones described in [42] that showed to minimize the contribution from any residual macroscopic anisotropy. Specifically, the measurements are performed in 8 planes, with their normals isotropically distributed on a sphere. Moreover, in each plane, the gradients with parallel directions (i.e. $\varphi = 0$) are rotated to point in 5 different directions. Thus, for each $\varphi$ value there are 40 measurements (8 planes × 5 in plane directions).

To assess the relative impact of exchange between soma and branches and between different branches, the signal angular dependence in the simulated cells is also compared with the analytical signal for a two-compartment model which accounts for restricted diffusion inside spheres and inside isotropically oriented finite cylinders.

*2.6.2 Mixing time dependence of the apparent microscopic anisotropy*

In the second DDE simulation, we study the behaviour of apparent µA in different synthetic cell configurations for various diffusion times and $\tau_m$. Towards this goal, we synthesize the signal from the well-studied DDE 5-design with 12 parallel and 60 orthogonal measurements [75] with three b-values ($b_4 = \{1, 2, 4\}$ ms/µm$^2$ for D = 2 µm$^2$/ms and $b_4 = \{4, 8, 16\}$ ms/µm$^2$ for D = 0.5 µm$^2$/ms), various diffusion times $\Delta_4 = \{5, 10, 20, 30, 45, 60, 80\}$ ms and $\tau_m$ between 0 and 200 ms, without exceeding a total sequence duration of 250 ms. Then, we calculate the apparent µA at each b-value based on the difference between measurements with parallel and orthogonal gradients [75].

Although we know from simulations that this metric slightly underestimates the expected microscopic anisotropy of the system [44] and at short $\tau_m$ it also reflects the effects of restriction size, it is a robust metric for comparing the trends between cells with different dendritic tree complexities.

Moreover, to investigate the effect of the exchange between spherical soma and branches and between different branches, we also compared the µA values from the simulations with those obtained from the analytical two-compartment model (sphere + isotropically oriented finite cylinders).

*2.7 Noise considerations*

To detect the effect of changing a certain parameter, such as spherical soma size or branching order, the differences incurred on the signal or on an estimated index (e.g. MD) should be larger than the variations due to noise. To study this effect, after the signal was averaged over the 10 cellular configurations, $N_{noise}$ = 1000 instances of Gaussian noise with standard deviation σ = 0.05 (i.e. corresponding to an SNR of 20 in the b=0 data) was added to each diffusion measurement. After adding noise, the signal is averaged over different directions, according to the protocols described in each experiment, followed by the computation of various indices (e.g. MD and MK). The differences incurred on the signal or on the estimated indices when changing a certain parameter are considered "detectable" if different from the noise induced variations with statistical significance assessed through two-tailed t-test and p<0.01. Further details are reported in section S2 of Supplementary Information.

## 3. Results

*3.1 Effect of soma size and branching order on SDE measurements*

*3.1.1 Spherical soma size can impact the b value dependence of the direction-averaged signal*

Figure 2 shows the signal b-value dependence for a diffusion time of 80 ms, for synthetic cells with different branching orders and spherical soma size. Focusing on soma, we find that soma size impacts the high b value dependence of the normalized direction-averaged signal, for both simulated water (Figure 2a) and metabolite (Figure 2b) diffusivities. In particular, for b>5 ms/μm$^2$, we observe a curvature (convexity) of the signal as a function of b value which increases when soma size increases. This effect is more clearly shown in Figure 3a). This effect is even more pronounced for the shorter diffusion times of 10 and 30 ms, where a difference in signal between different soma diameters occurs at lower b values, as illustrated in Figure S2. However, if the diffusion time is very short relative to the restriction sizes, the effect of different soma sizes is reduced. This can be seen for instance in Figure S1b, where for Δ = 10 ms and D = 0.5 μm$^2$/ms the mean squared root displacement $\sqrt{6D\Delta}$ = 5.4 μm is smaller than the soma diameters.

*Noise considerations:* Figures S5a) and b) in Supplementary Information present the signal differences between cell configurations with $D_s$ = 8 μm and larger diameters as a function of b-value for diffusion times of 80 and 10 ms, respectively. For Δ = 80 ms, the signal differences are largest for high b values (3 – 20 ms/μm$^2$) both for high and low diffusivity values. For shorter diffusion times, the differences shift towards lower b-values and the noisy curves become separable for b-values around 0.5 ms/μm$^2$ for D = 2 μm$^2$/ms and around 1 ms/μm$^2$ for D = 0.5 μm$^2$/ms.

*3.1.2 Cylindrical branch diameter can impact the high b value dependence of the direction-averaged signal at short diffusion times*

As explained in section 2.1 and 2.2, to keep the cellular volume fraction occupied by the soma constant, we adapted the $D_b$ value keeping fixed L and $N_{proj}$ according to Eq. [2]. For most substrates, which have projections with diameters ≤ 1.5 μm, the branch diameter does not have an impact on the b value dependence of the signal. Nevertheless, for the substrates which have a larger branch diameter (i.e. cells with low $N_b$ and large soma size), the signal decay curves at short diffusion time (Δ = 10 ms) start diverging at very large b values (b > 10 ms/μm$^2$). This effect is especially clear for simulations with D = 0.5 μm$^2$/ms, as illustrated in Figure S1b).

*Noise considerations:* The effect of branch diameter is detectable in the presence of noise, especially for D = 0.5 μm$^2$/ms, as illustrated in Figure S6a) which shows large differences between cells with different soma sizes (and implicitly different branch diameters) at high b values.

*3.1.3 Negligible impact of branching order of cylindrical projections on the b value dependence of the direction-averaged signal*

As illustrated in Figure 2, the signal decay curves as a function of b-value are similar for synthetic cells with different branching orders for the SDE sequence parameters chosen in these simulations, i.e. short gradient duration and diffusion time up to 80 ms. Moreover, comparing the signal decay from Monte Carlo simulations with a theoretical model of two, non-exchanging, compartments (Figure 3b) also shows a good agreement between the curves, especially for $N_b$ = 6, implying that for the SDE sequences investigated here, the b-value dependence of the signal cannot directly inform on the complexity of the cell dendritic tree, expressed in terms of branching order $N_b$. Even better agreement is observed for the shorter diffusion times of 10 and 30 ms.

*Noise considerations:* Indeed, the results presented in Figures S5b) for Δ = 80 ms and in S6b) for Δ = 10 ms show that for cells with $N_b$ = {4 ,6} there are just very small differences between the simulated signal and the theoretical compartment model, and for the larger spherical soma sizes the shaded area describing the standard deviation of the noise overlaps with the theoretical model. The differences are even smaller for the lower diffusivity value $D = 0.5$ µm$^2$/ms.

*3.1.4. Soma-branch exchange can impact the b value dependence of the direction-averaged signal at long diffusion times*

For structures with $N_b = 1$ (which have the largest potential of soma-branches exchange, see $p_{ex}$ values in Table S1), the simulated signal is below the theoretical curve for a range of intermediate b values (1-10 ms/µm$^2$), especially for $D = 2$ µm$^2$/ms and $\Delta = 80$ ms. This effect is less pronounced for the shorter diffusion times and is not present for the structures with branched projections and implicitly lower $p_{ex}$ values, pointing out this effect rises from the exchange between spherical soma and cylindrical projections.

*Noise considerations:* As illustrated in Figures S5b) and S6b), this effect is more pronounced for $\Delta = 80$ ms, and significant differences between theoretical and simulated curves are present for both $D = 2$ µm$^2$/ms and $D = 0.5$ µm$^2$/ms, given the simulated SNR of 20.

*3.1.5 Spherical soma size can impact the MD and MK time dependence at short to intermediate diffusion times*

Figure 4 and Figure 6 illustrate the MD and MK time dependence for large and small cell domains with various spherical soma diameters and branching orders, for diffusivities mimicking both water diffusion ($D = 2$ µm$^2$/ms) and metabolites diffusion ($D = 0.5$ µm$^2$/ms).

For all substrates, the spherical soma size has a marked influence on the time dependence of MD and MK, as also illustrated in Figure 5a) and 7a). For smaller spherical soma sizes, there is a sharp decay in MD and increase in MK at short diffusion times (~ 10 ms), followed by a slower change at longer diffusion times. For the larger spherical soma sizes this regime extends up to ~100 ms. These patterns are further shifted to longer diffusion times for the simulations with low diffusivity as illustrated in Figure 5a) and 7a).

*Noise considerations:* Figure S7 and S8 from Supplementary Information present the differences in MD and MK between cells with $D_s = 8$ µm and cells with larger diameter. For intermediate diffusion times, the differences in MD and MK are larger than the variations due to noise, both for $D = 2$ µm$^2$/ms and $D = 0.5$ µm$^2$/ms.

*3.1.5 Cylindrical branch diameter can impact the MD and MK time dependence at short diffusion times*

It is worthwhile to note that in some cases reported in Figure 4 and Figure 6, we can additionally see the impact of non-negligible cylindrical branch diameter on the simulated MD and MK time dependence, especially at diffusion times <~20 ms.

For D = 2 µm$^2$/ms, in Figure 4a) and Figure 6a), the effect of finite branch diameter is negligible in most substrates, except for the smaller cells (L = 400 µm) with large soma sizes ($D_s$ = 16, 20 µm) and $N_b$ = 1 and 2, where the branch diameter is larger than 1.9 µm. In these cases, we see a higher MD and a lower MK at the very short diffusion times[64, 65]. Consequently, for smaller cell domain and lower $N_b$, $D_b$ values are large enough to have a non-negligible impact on the MD and MK diffusion time dependence at relatively short diffusion time.

For D = 0.5 µm$^2$/ms, in Figure 4b) and Figure 6b), the effect of branch diameter on MD and MK is more pronounced, as smaller diameter values are detectable as expected from previous analyses [64, 65].

*Noise considerations:* The effect of cylindrical branch diameter is also clearly illustrated in Figure S7a) and S8a), where signal differences due to an increase in branch diameter are also detectable in the presence of noise for short diffusion times.

*3.1.7 Small impact of branching order of cylindrical projections on MD (but not MK) time dependence at long diffusion time*

For all substrates, Figure 4 and Figure 6 show that there is a visible departure in MD and MK between the cells with different branching orders for long diffusion times (> 100 ms), which is more pronounced for longer diffusion times and smaller cells. These differences in MD and MK values are determined by two effects, namely the cell branching itself and a difference in exchange probabilities between soma and branches, as illustrated in Table S1.

To separate these effects, Figure 5b) and Figure 7b) are comparing the signal decay from Monte Carlo simulations with a theoretical model of two, non-exchanging, compartments, namely a sphere and isotropically oriented capped cylinders, with the same effective diameters as the values from Table 1. To investigate the effect of branching we focus on substrates with $N_b = \{4, 6\}$ which have a low potential of exchange between soma and branches.

For substrates with larger $N_b$ values, and implicitly lower exchange potential, we see a good alignment between the simulated and theoretical curves, both for MD and MK. For the most branched cells ($N_b = 6$), we see slightly lower MD values compared to the theoretical curves for very long diffusion times > 200 ms which is due to the cell projections' branching, while the MK values remain lower than the theoretical values for all $N_b$ investigated here. At long diffusion times, the overall domain size also has an impact on the MD and MK values, which have a more pronounced change with diffusion time for smaller cells with L = 400 μm compared to cells with L = 1000 μm, a trend also captured by the finite cylinders from the two-compartment model as illustrated in Figure 5b) and Figure 7b).

*Noise considerations:* To better assess the detectability of these effects, Figure S7b) and S8b) plot the signal differences between the simulated signal, which includes the effects of branched projections ($N_b > 1$) as well as the exchange between soma and branches, and the theoretical signal for a compartment model which does not include these effects, in the presence of noise. For substrates with $N_b = \{4, 6\}$, the impact of branching order on MD is small, nevertheless, for cells with $N_b = 6$ at long diffusion times, there is a small but significant decrease in MD compared to the theoretical model. When considering MK, for structures with $N_b = 6$, the shaded areas, showing the standard deviation of MK for an SNR of 20, overlap with the theoretical values.

*3.1.8 Soma-branch exchange can impact the MD and MK time dependence at long diffusion times*

For cells with low branching orders $N_b = \{1,2\}$ we see higher MD values and lower MK values in the simulated cells compared to the theoretical model for diffusion times larger than 20-30 ms. This trend was seen for other values of the soma diameter as well and is due to the exchange between the soma and the branches which is the most pronounced for cells with $N_b = 1$ compared to other cells due to the larger branch diameter (see $p_{ex}$ values in Table S1).

*Noise considerations:* For diffusion times above ~100 ms, we see a significant increase in MD and decrease in MK due to exchange between soma and branches for cells with $N_b = \{1, 2\}$, for the considered SNR of 20.

*3.1.9 Distinct regimes for the impact of cylindrical branch size, spherical soma size, soma-branch exchange and branch-branch exchange*

The results in the previous sections suggest the existence of distinct regimes where the effects of cylindrical branch diameter, spherical soma size, soma-branch exchange and the branch-branch exchange due to branching of cellular projections (here parametrized by $N_b$), dominate the MD and MK time dependence. Specifically, for water diffusion:

a) at very short diffusion times, i.e. ≤ 10 ms, branch diameters ( > 1.5 μm) seem to drive the MD and MK time dependence;
b) at short to medium diffusion times, i.e. ~10-50 ms, the spherical soma diameter $D_s$ seems to drive the MD and MK diffusion time dependence;
c) at medium to long diffusion times, i.e. ≥ 50 ms, the exchange between soma and branches seems to impact the MD and MK diffusion time dependence (Figure 3),
d) at very long diffusion times, i.e. > 200 ms, the branch-branch exchange has a dominant effect on MD.

These regimes seem to hold for metabolites too, once longer diffusion times are considered to compensate for the much slower diffusion coefficient of metabolites compared to water (Figure 4b), Figure 6b), Figure S6 and Figure S7). Also, it is worth noting that the soma-branch exchange has an opposite effect to the branch-branch exchange on the MD and MK time dependence.

*3.2 Effect of soma size and branching order on DDE measurements*

*3.2.1 Spherical soma size can impact the DDE signal's angular modulation at short mixing times*

Figure 8a) plots the angular DDE modulation as a function of mixing time for cells with different spherical soma size for $N_b = 1$ and $N_b = 6$ and a cell domain L = 400 μm, for the water diffusivity D = 2 μm²/ms. For the DDE sequences with short diffusion time (Δ = 5 ms) considered in this simulation, spherical soma size has only a small effect on the angular modulation, shown by the bell-shaped curve for $D_s = 8$ μm at $\tau_m = 1$ ms. For the same cellular configurations with $D_s = 8$ μm, the signal is slightly shifted compared to the other soma sizes, nevertheless all curves follow the same amplitude modulations for $\tau_m$ >10 ms. The shift occurs due to a more pronounced effect of restricted diffusion inside the soma, which is closer to the mean squared displacement of $\sqrt{6D\Delta}$ ~ 8 μm compared to the larger soma diameters, for which only part of the spins will probe the boundary during the given diffusion time. The larger effect of restricted diffusion for $D_s = 8$ μm can also be seen from the higher difference between the DDE with parallel (0º) and orthogonal (180º) gradient orientations. For these DDE sequences, the effect of spherical soma size for metabolites diffusivity $D = 0.5$ μm$^2$/ms is negligible, as it is clearly illustrated in Figure S4a) for structures with $N_b = 6$ where the effect of branch diameter is negligible, and all curves overlap. Moreover, for the shortest mixing time, there is no difference between parallel and anti-parallel signals, which is also a signature of restriction.

*Noise considerations:* When considering the difference between DDE measurements with parallel and anti-parallel gradients at $\tau_m = 1$ ms as a probe of restriction size, for the larger diffusivity value $D = 2$ μm$^2$/ms, the signal differences are ~ 0.07 for cells with $D_s = 8$ μm, ~ 0.02 for $D_s = 12$ μm and below 0.01 for the largest diameters. Thus, even for the short diffusion time used in these simulations, the impact of the smaller soma sizes on the DDE signal at short mixing time is above the noise level, for the noise distribution considered in this study with $\sigma = 0.05$ and averaging over the 40 directions.

*3.2.2 Cylindrical branch diameter can impact the DDE signal's angular modulation at short mixing times*

As illustrated in Figure S4, for simulations with $D = 0.5$ μm$^2$/ms and structures with branch diameter ≥ 1.5 μm, there is a clear effect both on the DDE signal values and on the mixing time dependence of the angular modulation shape, which at very short mixing times has the characteristic bell-shaped curve for restricted diffusion. Also, for the substrates with the larger branch diameters, the overall amplitude of the angular modulation at long mixing times is lower than for substrates with smaller branch diameters, nevertheless it does not vary with mixing time once the long-time regime with respect to the diameter values has been reached.

*Noise considerations:* When considering the difference between DDE measurements with parallel and anti-parallel gradients at $\tau_m = 1$ ms as a probe of restriction size, for $D = 0.5$ μm$^2$/ms the signal differences are ~ 0.03 for cells with $D_b \sim 2$ μm, which is above the noise level for the distribution considered in this study with $\sigma = 0.05$ and averaging over the 40 directions.

*3.2.3 Branching order of cylindrical cellular projections can impact the mixing time dependence of the DDE signal's angular modulation*

Figure 8b) presents the angular DDE modulation as a function of mixing time for synthetic cells with different branching orders for $D_s$ = 8 μm and $D_s$ = 20 μm and a cell domain L = 400 μm. The plots show a decrease in the amplitude of the DDE signal modulation with mixing time for the cells with branched projection, and the decrease is larger for larger values of $N_b$. To further investigate this effect, Figure 9 compares the simulated angular DDE modulation with the signal provided by a non-exchanging two-compartment model consisting of diffusion restricted in a sphere and finite isotropically oriented cylinders with the parameters described in Table S1, which is exemplified for $D_s$ = 8 μm. Figure 9a) shows a good agreement between simulated and theoretical curves for cells with straight projections ($N_b$ = 1) for the entire range of mixing times, while Figure 9b) indeed shows a decrease in the modulation amplitude with mixing time for the branched cells with $N_b$ = 6, both for D = 2 and 0.5 μm$^2$/ms. Similar trends have been observed for other substrates and soma diameters. For simulations with D = 0.5 μm$^2$/ms or for larger cell domains (L = 1000 μm), the decrease is less pronounced compared to the results for D = 2 μm$^2$/ms, as less spins travel from one segment to the other in the same time interval. The results are shown for the sequences with b = 4 ms/μm$^2$ for D = 2 μm$^2$/ms and b = 16 ms/μm$^2$ for D = 0.5 μm$^2$/ms nevertheless, similar trends are seen for other b values.

*Noise considerations:* Analysing the difference between DDE measurements with parallel and orthogonal gradient orientations and how it changes with mixing time, Figure S9a) shows that for cells with highly branched projections ($N_b$ = 4 and 6), the amplitude modulation decreases with mixing time. The change in amplitude modulation between short mixing times and longer mixing times can be detected at $\tau_m$= 200 ms also when considering noisy data, both for D = 2 and 0.5 μm$^2$/ms. The decrease is less pronounced (< 0.015) for cells with straight projections, and the signals overlap within their standard deviations. Similar trends and detectability levels are observed when comparing the simulated DDE signal with the compartment model (Figure S9b).

*3.2.4 Negligible impact of soma-branch exchange on the mixing time dependence of the DDE signal's angular modulation*

For the DDE sequences investigated in this work, we see no direct effect of exchange between spherical soma and cylindrical projections, as the amplitude of the angular modulation for cells with $N_b$ = 1 (which have the highest exchange potential) does not vary with mixing time and closely matches the theoretical two compartment model, as illustrated in Figure 9 and Figure S9.

*3.2.5 Spherical soma size can impact the mixing time dependence of the μA for medium diffusion times*

To further investigate the effect of spherical soma size and branching order on DDE signal, Figure 10 presents the mixing time dependence of the apparent microscopic anisotropy for DDE sequences with short (a,c,e,g) and long (b,d,f,h) diffusion times, for cells with different branching orders, for D = 2 µm$^2$/ms and D = 0.5 µm$^2$/ms, respectively. The plots also compare the simulated values with predictions from a non-exchanging two-compartment model.

For DDE sequences with short diffusion time (Δ = 5 ms) we see similar µA trends for small and large soma values with D$_s$ = 8 µm and 16 µm, respectively (Figure 10a, 10c, 10e, 10g). For DDE sequences with longer diffusion time (Δ = 30 ms) we see an initial increase in µA with mixing time, which becomes more pronounced as the soma size increases, especially for the larger diffusivity value D = 2 µm$^2$/ms. For the smaller diffusivity value, i.e. D = 0.5 µm$^2$/ms, and larger soma diameter D$_s$ = 16 µm, the diffusion time of Δ = 30 ms is not long enough to probe the spherical restriction and the initial increase with mixing time is no longer apparent.

*3.2.6 Cylindrical branch diameter can impact the mixing time dependence of the µA for short diffusion times*

Besides the soma size effect, the plot in Figure 10g (D = 0.5 µm$^2$/ms, Δ = 5 ms and large soma size D$_s$ = 16 µm) also shows the effect of finite branch diameter which is reflected by different plateau values of the theoretical curves corresponding to cells with different N$_b$ values, which by design, have different branch diameters. As illustrated in Figure 9g), for sequences with short diffusion time Δ = 5 ms and small diffusivity D = 0.5 µm$^2$/ms, the effect of branch diameters > ~ 1.5 µm can be observed on the µA values, which are different for different substrates, as also reflected by the theoretical model. For the synthetic cells with the largest branch diameter, there is also a sharp increase in µA at very short mixing times (data not shown).

*Noise considerations:* The effects of branch diameters > ~ 1.5 µm can be detected for the metabolites' diffusivity D = 0.5 µm$^2$/ms, following a similar rationale to 3.2.2.

*3.2.7 Branching order can impact the mixing time dependence of the µA*

Figure 10 shows lower values of microscopic anisotropy for cells with branched projections (i.e. N$_b$ = 4 and N$_b$ = 6) compared to the values obtained for cells with straight projections (N$_b$ = 1) as well as the corresponding theoretical curves from the two-compartment model. Moreover, for DDE sequences with short diffusion time (5 ms), we see a decrease in μA with mixing time as the branching order increases. This effect can be seen for higher and lower diffusivity values, both for cells with a small soma diameter of 8 μm, as well as a larger soma diameter of 16 μm (Figure 10a,10c, 10e and 10g). This decrease in μA cannot be captured by the analytical two-compartment model, which, apart from a slight increase at short $\tau_m$ in the case of $D_s$ = 8 μm, shows no μA dependence on mixing time.

For the sequences with longer diffusion time of 30 ms, the mixing time dependence of μA is dominated by soma size and the effect of branching is less pronounced compared to the DDE sequences with short diffusion times, as illustrated in Figure 10b,d for D = 2 μm$^2$/ms and in Figure 10f, h for D = 0.5 μm$^2$/ms, respectively. In this case, the difference between cells with $N_b$ = 1 and $N_b$ = 6 is mainly reflected by an overall shift in the μA values, rather than a different dependence (i.e. a different slope) on $\tau_m$. Similar trends to the ones presented in Figure 10 are observed for other cellular configurations as well.

*Noise considerations:* As illustrated in Figure S10a), the detectability of μA$^2$ differences for noisy DDE with Δ = 5 ms is similar to the results presented in section 3.2.2. Branched cells ($N_b$ = 4 and 6) show a decrease in μA$^2$ with mixing time which can be detected considering the conditions of this analysis, i.e. 12 parallel and 60 perpendicular directions, noise standard deviation of 0.05 for each measurement. The decrease is significant both for D = 2 μm$^2$/ms and D = 0.5 μm$^2$/ms. For sequences with Δ = 30 ms (Figure S10b), the decrease in μA$^2$ with mixing time is less pronounced, nevertheless, the branched cells show a larger mismatch between the theoretical and the simulated data, compared to the cells with straight projections.

*3.2.8 Negligible impact of soma-branch exchange on the mixing time dependence of the μA*

For the DDE sequences investigated in this work, we see no direct effect of exchange between spherical soma and cylindrical projections, as μA values estimated for cells with $N_b$ = 1 (which have the highest exchange potential) do not vary with mixing time and closely match the theoretical two compartment model, as illustrated in Figure 10 and Figure S10.

## 3.3 Summary of results

A summary of the results for the sequences employed in this work is presented in Table 1.

| Diffusion sequence | Measurement | Intracellular probe molecule | Sensitivity to spherical soma size | Sensitivity to branching of cylindrical projections (branch-branch exchange) | Sensitivity to cylindrical projections size | Sensitivity to soma-branch exchange |
|---|---|---|---|---|---|---|
| SDE (short $\delta = 1$ ms) | Signal intensity at fixed diffusion time and varying gradient strength | Water $D_0 = 2$ μm²/ms | **YES** for b>3 ms/μm² and $\Delta \leq 30$ ms | **NO** | **NO** because it requires higher b values than those simulated here | **YES** for $0.5 < b < 10$ ms/μm² and $\Delta \geq 30$ ms NOTE intra-extracellular exchange may mask the effect |
| | | Metabolites $D_0 = 0.5$ μm²/ms | **YES** for b>10 ms/μm² and $\Delta \leq 100$ ms | **NO** | **YES** for b>10 ms/μm², $\Delta \leq 10$ ms and diameters $\geq 1.5$ μm | **YES** for $2 < b < 10$ ms/μm² and $\Delta \geq 80$ ms |
| SDE (short $\delta = 1$ ms) | MD at varying diffusion time | Water $D_0 = 2$ μm²/ms | **YES** for $\Delta \in [5\text{-}40]$ ms | **YES** for $\Delta > 200$ ms NOTE intra-extracellular exchange may mask the effect | **NO** | **YES** for $\Delta \geq 20$ ms NOTE intra-extracellular exchange may mask the effect |
| | | Metabolites $D_0 = 0.5$ μm²/ms | **YES** for $\Delta \in [40\text{-}200]$ ms | **YES** for $\Delta > 1000$ ms | **YES** for $\Delta \leq 10$ ms and diameters $\geq 1.5$ μm | **YES** and $\Delta \geq 100$ ms |
| SDE (short $\delta = 1$ ms) | MK at varying diffusion time | Water $D_0 = 2$ μm²/ms | **YES** for $\Delta \in [20\text{-}100]$ ms | **NO** | **NO** | **YES** for $\Delta \geq 20$ ms |
| | | Metabolites $D_0 = 0.5$ μm²/ms | **YES** for $\Delta \in [80\text{-}400]$ ms | **NO** | **YES** for $\Delta \leq 10$ ms and diameters $\geq 1.5$ μm | **YES** and $\Delta \geq 100$ ms |
| DDE (short $\delta = 1$ ms) | Amplitude of angular signal modulation, $\Delta = 5$ ms | Water $D_0 = 2$ μm²/ms b = 4 ms/μm² | **YES** for $t_m \to 0$ and small soma sizes, as larger sizes require longer diffusion times | **YES** by contrasting measurements with $t_m \sim 1$ ms and $t_m > 20$ ms | **NO** | **NO** |
| | | Metabolites $D_0 = 0.5$ μm²/ms b = 16 ms/μm² | **NO** for $t_m \to 0$ because it requires longer diffusion times than simulated here | **YES** by contrasting measurements with $t_m \sim 5$ ms and $t_m > 80$ ms | **YES** for $t_m \to 0$ and diameters $\geq 1.5$ μm | **NO** |
| DDE (short $\delta = 1$ ms) | μA | Water $D_0 = 2$ μm²/ms b = 4 ms/μm² | **YES** for $t_m \to 0$ and $\Delta > 5$ ms | **YES** by contrasting measurements with $t_m \sim 1$ ms and $t_m > 30$ ms | **NO** | **NO** |
| | | Metabolites $D_0 = 0.5$ μm²/ms b = 4 ms/μm² | **YES** for $t_m \to 0$ and $\Delta > 20$ ms | **YES** by contrasting measurements with $t_m \sim 5$ ms and $t_m > 100$ ms | **YES** for $t_m \to 0$, $\Delta \sim 5$ ms and diameters $\geq 1.5$ μm | **NO** |

*Table 1 A summary of different dMRI/dMRS regimes and acquisitions where the measurements show sensitivity to different substrate properties, namely spherical soma size, branching of cylindrical projections, cylindrical projection size, soma-branch exchange.*

## 4. Discussion

This work employs MC simulations of diffusion within realistically connected neuron-like meshes in order to study the effect of branching order of cellular projections and soma size on the signal measured using both single (SDE) and double (DDE) diffusion encodings. We investigate both fast and slow diffusion, to mimic intracellular water and metabolites diffusion, respectively. Although we see only a small effect of branching on the signal measured with standard SDE sequences, our key results show that DDE acquisitions with variable mixing time could provide information about the branching order of the cellular projections (Table 1). Moreover, we show that as the cellular projections become more branched, there is a more pronounced decrease in measured apparent µA with mixing time, which could be used as a signature of this feature.

The SDE results presented here support the use of simpler compartmental models for disentangling the soma contribution to the overall signal, for example SANDI [28], as the branching makes little difference on the time and b-value dependence for values typically used in diffusion experiments. Nevertheless, this work also points at the necessity to carefully choose the experimental design, since the impact of soma seems to dominate the signal within a specific time and b-value window (Table 1). These results are also in line with recent preliminary data showing that branching has little impact on the b-value power law characteristic for straight cylinders with infinitesimally small radius [76].

*4.1 Comparison with metabolites dMRS literature*

Our simulations are focused on the intracellular space only and therefore are directly relevant for metabolites dMRS measurements. Our results at slow diffusivity (i.e. mimicking metabolites diffusion) are in good agreement with experimental evidences from previously published SDE [30, 32, 34, 77-88] and DDE [33, 89-92] measurements of metabolites diffusion through dMRS.

To our knowledge, there are a few studies investigating the metabolites MD time-dependence but none investigating the MK time-dependence. Our simulation results match the observed MD decrease at increasing diffusion time for the mostly intracellular metabolites such as N-acetylaspartate (NAA), Creatine (Cr), Myo-Inositol (Ins) and Choline (Cho) [30, 32, 34, 77-88]. To compare more directly our simulation results with experimental findings, we have reanalyzed the data from [110] and estimated metabolites MD and MK at Δ ~ 64 and 254 ms to be on average: MD ~ 0.115 and 0.085 $\mu m^2$/ms; MK ~ 1.45 and 1.75 (see Figure S11 in Supplementary Information). Our simulation results for spherical soma of 12 μm in diameter suggest MD ~ 0.120 and 0.090 $\mu m^2$/ms; MK ~ 1.40 and 2.20, so in agreement with the experimental findings for purely intracellular metabolites. As a novel result, our simulations of MK time-dependence suggest that MK time-dependent measurements of intracellular metabolites can be sensitive to the exchange between soma and projections, pointing at the interesting possibility to design metabolites SDE acquisitions optimized to measure this exchange mechanism.

Concerning DDE acquisitions, measuring the diffusion of purely intracellular metabolites in-vivo in rat brain, Shemesh et al. [89, 90] showed that DDE signal from NAA and Ins displayed characteristic amplitude modulations reporting on confinements in otherwise randomly oriented anisotropic microstructures for both metabolites. More recently, Vincent et al. [91] showed in vivo in mouse brain that a simple geometrical model of randomly oriented cylinders is not able to accurately explain the experimental DDE data, and that a more complex model incorporating branching (and/or other subtle structures such as spines) is indeed needed. Our simulation results on the amplitude of angular signal modulation from DDE measurements using brain metabolites diffusivity match these experimental observations, further supporting the interpretation that indeed branching of complex neural cell structures significantly impacts intracellular metabolites diffusion under the experimental conditions investigated by Shemesh et al. [89, 90] and Vincent et al. [91]. Moreover, the recent DDE measurements of metabolites diffusion in vivo in human brain by Lundell et al. [92] corroborate these results and additionally show that the microscopic fractional anisotropy of tCho, intracellular metabolite preferentially found in glial cells, is significantly lower in gray matter than in white matter. The authors speculated that a possible explanation is that in gray matter a significant fraction of tCho may be found in protoplasmic astrocytes. These astrocytes, found extensively in human gray matter, are highly branched cells, significantly more so than their fibrous counterparts in white matter. Our simulation results support this explanation, showing that indeed higher branching can reduce the DDE signal's amplitude angular modulation and the estimates of derived apparent microscopic anisotropy.

*4.2 Comparison with water dMRI literature*

A few studies investigated MD and MK time-dependence in GM through water dMRI [93-96]. They reported a marked decrease of MD and increase of MK as diffusion time increases up to ~20 ms, then the decrease of MD is much less pronounced, and MK starts decreasing with increasing diffusion times. Our simulation results match the observed decrease of MD and increase of MK at short diffusion times (i.e. Δ ≤ 20 ms) [93-95] but do not match the observed behaviours of MD and MK at longer diffusion times (i.e. Δ > 20 ms) [96]. As pointed in the literature [94, 96], one of the possible mechanisms explaining the observed MD and MK time-dependence at long diffusion times in GM could be the intra-extracellular exchange. Since we have not accounted for this mechanism in our simulations, we hypothesize that the discrepancy of our simulation results with respect to published experimental data at diffusion times longer than ~20 ms can be due to intra-extracellular exchange. This indirectly suggests that intra-extracellular exchange in GM may affect the measured MD and MK time-dependence for diffusion times > 20 ms more than the soma-branch exchange and the branch-branch exchange (both included in our simulations).

The conclusions concerning water DDE measurements may also be altered by the exchange with extracellular space. Earlier studies [27, 97] exclusively based on water dMRI measurements at b values ≤ 2 ms/μm$^2$ showed lower μA in GM. Given the relatively low b value, this could be explained by fast intra-extracellular exchange as well as significant contribution from mostly isotropic extracellular space. However, the recent study by Lundell et al. [92] compared DDE measurements of metabolites and water diffusion in healthy human brain and showed that water μA at high b values (> 2 ms/μm$^2$) in GM is similar to that of some purely intracellular metabolites (e.g. NAA), suggesting that the signal from the extracellular space is effectively suppressed at high b values (~ 4-7 ms/μm$^2$) and that intra-extracellular exchange has small effect under the investigated experimental conditions (gradient separation ~45 ms; mixing time ~5 ms). Our simulation setup includes these experimental conditions and indeed our results mirrors the experimental observations by Lundell et al. for both water and metabolites. The DDE simulations show similar trends for different b-values, nevertheless for water this may change at low b-values (~1 ms/μm$^2$) due to the effect of extracellular space.

Moreover, tortuosity values in the cytoplasm for water and NAA were found remarkably similar, while exhibiting a clear difference between gray and white matter, suggesting a more complex cytomorphology of neuronal cell bodies and branching dendrites in GM compared to WM [92]. These findings are also in good agreement with our DDE results, which support the hypothesis that the more complex cytomorphology of neural cells in GM can significantly impact DDE measurements and DDE acquisitions can be potentially used to quantify it.

Finally, a study by Ianus et al [44] measuring the time and frequency dependence of μA in mouse brain, ex-vivo, showed that the observed μA time dependence cannot be explained by any simple model of water restricted in cylinders and/or spheres, also supporting our simulation results at the fastest diffusivity (i.e. mimicking intracellular water diffusion) and the possibility of measuring new features of brain cell morphology, such as branching.

*4.3 Impact of soma-branch exchange and branch-branch exchange*

In this work we investigate two potential mechanisms of exchange in neural cells that have been previously ignored: the soma-branch and the branch-branch exchange. Our results show that soma-branch exchange can have a significant effect on SDE measurements within an experimental regime which is quite different from that of the branch-branch exchange (see Table 1). This suggests that it may be possible to tune dMRI/dMRS measurements to disentangle the two mechanisms and potentially quantify them. For metabolites dMRS, these conclusions directly point towards exciting new perspectives for future experiments, informing the design of acquisitions aiming to disentangle and measure these two exchange mechanisms. Perhaps, more sophisticated encodings of the diffusion gradients [46, 98] could be envisioned, also benefitting from filtering techniques such as the relaxation-enhanced dMRS at ultrahigh field [89]. Such measurements could be important in the context of measuring in vivo brain plasticity [3] in psychiatric disorders [99] or in the context of degenerative diseases [100], providing new insights into changes in neural cell soma-branch connectivity or branching order of projections.

Concerning water dMRI, an additional third mechanism of exchange can play a significant role: the intra-extracellular exchange. Based on the comparison of our results on water MK time-dependence with recent experimental findings (see previous section), we can hypothesis that the intra-extracellular exchange may have a dominant impact at diffusion times $> 20$ ms, hence potentially outweigh the influence of branch-branch exchange. However, whether the intra-extracellular exchange actually dominates over the other two mechanisms of exchange is a new open question for future works.

*4.4 Potential impact of extracellular signal and intra-extracellular exchange on water dMRI results*

In this work we focus on intracellular signal only, in order to systematically investigate the effect of branching order of cellular projections and soma size, without considering exchange with any extracellular space. The exchange between intra and extracellular space could impact water dMRI experiments and might alter some of the corresponding conclusions. Therefore, the prominent direct application of our results is in intracellular metabolite dMRS studies, with our simulations mimicking water diffusion surely of interest for dMRI applications, when the effect of water exchange can be considered negligible. For example, as already discussed in section 4.2, intra-extracellular exchange could alter our results on the MD and MK time dependence, potentially explaining the mismatch of our simulations with the experimentally observed behaviours of MD and MK at medium-long diffusion times ( > 20 ms). However, it is still unclear how fast the intra-extracellular water exchange is in vivo in brain grey matter and at which time scale its effect becomes significant. While substantial information exists on water exchange through cellular membranes in vitro, the in vivo information remains limited and controversial. From experiments using in vitro cultures of rat cortex, Bai et al. [101] and Yang et al. [102] consistently estimated the apparent water exchange time to be of ~ 0.5-0.8 seconds. On the other hand, studies using a technique called filter-exchange imaging (FEXI) [103] consistently measured apparent water exchange time in vivo in human brain cortex of ~1.4-2.5 seconds [104-106]. According to these FEXI estimates, for in vivo dMRI applications we can consider water exchange effects to be negligible in brain grey matter for diffusion and mixing times much shorter than ~2 seconds. This is the case concerning our DDE results, where the longest mixing time is 0.2 seconds, while it may indeed affect our conclusions about the MD and MK time-dependence from SDE measurements, where the longest diffusion time is 2.450 seconds. More quantitatively, assuming the FEXI framework and parameters previously reported in the literature for in vivo human brain grey matter [104-106], we estimate that the signal difference between the DDE signal (total b value = 4 ms/μm$^2$) at $\tau_m$ = 1 ms and $\tau_m$ = 200 ms due to exchange would be ~5 times smaller than the signal amplitude difference due to the branching of cellular projections as quantified in Supplementary Information section S3 and Figure S7. Of course, these considerations should be revised by appropriate rescaling if new experimental evidences would suggest faster apparent water exchange times in gray matter. Moreover, we note that highly permeable cellular projections could still support long voxel-level exchange times, if the soma and myelinated axons had a low permeability. Nevertheless, recent findings using metabolites and water DDE in humans [92] show high microscopic anisotropy in grey matter measured at high b-values and challenge previous results in [27, 97], pointing towards the possibility of negligible impact from extracellular signal and intra-extracellular exchange on DDE measurements under specific experimental conditions, which have been investigated here as well. Future works are needed to assess when and to what extent the intra-extracellular exchange and different permeability for different cellular sub-compartments (e.g. nucleus, soma, projections, myelinated axons etc) could change our results and conclusions.

*4.5 Limitations and future works*

One main limitation of this simulation study is that it does not include exchange with extracellular space, which can also affect MD and MK time dependence for SDE sequences as well as the mixing time dependence of angular DDE signals and microscopic anisotropy metrics in case of water-based measurements, especially if longer diffusion / mixing times are considered. We chose to focus on intracellular signal only in order to systematically investigate the effect of soma size and branching order of cellular projections, without the added complexity of cellular packing and exchange between intra and extracellular spaces, which is a research topic on its own. Furthermore, there is currently lack of computational tools able to densely pack complex cellular structures like those considered in this work into realistic virtual tissues. This is a crucial aspect necessary to assess any realistic and sensible impact of exchange with brain extracellular space. Moreover, the meshes simplify the cellular structure and do not account for cellular nuclei or axonal features such as variations in calibers and axonal undulations that can have a larger impact on the dMRI signal than the cylinder diameter itself [107, 108]. However, it is worthwhile to note that substantial steps forward have been recently achieved for white matter numerical phantoms [53, 56, 109]. Future works will aim at adapting these kinds of approaches for realistic grey matter numerical phantoms generation, enabling an exhaustive study of also the intra-extracellular exchange. Nevertheless, the results presented here for $D = 0.5$ μm$^2$/ms, which show similar SDE and DDE signatures, are highly relevant for spectroscopy studies which investigate intracellular metabolites, as proven in [33, 89-91], and the use of potentially high b vales (b>4 ms/μm$^2$) may mitigate the effect of extra-cellular water in dMRI applications.

The meshes used in these simulations have been designed in order to assess the effect of branching and soma size on the diffusion time and b-value dependence in as fair a way as possible. As the signal fraction of restricted diffusion has a great influence on the time-dependence of the measured signal, we designed the meshes to have the same cellular volume fraction occupied by the soma (~30%). To achieve this, we have adjusted the diameter of the branches. Therefore, cells with $N_b = 6$, have very thin projections, and thus a smaller exchange potential between the soma and the branches. On the other hand, cells with no branching ($N_b = 1$) have wider projections, with diameter >1.5 μm in some cases, thus the time dependence of restricted diffusion inside the cylinders can become noticeable at short diffusion times, especially for $D = 0.5$ μm$^2$/ms. Moreover, for the same cells with larger diameters, the effects of exchange between the branches and the soma can also play a role for medium to long diffusion times. As the focus of this work was to investigate the effect of spherical soma size and branching order of projections, the sensitivity to branch diameter is a secondary effect given by the way the substrates were designed, and not a sought-after contrast. As described in previous literature [64, 65], other sequences might significantly improve the contrast to the diameter of cylindrical objects, which is outside the scope of this work.

These simulations have also considered ideal sequences with short gradient duration δ = 1 ms and covered a wide range of b-values and diffusion times, leading to very high gradient strengths for some configurations, not all of which are feasible in practice. For instance, the maximum gradient strength to achieve SDE sequences with $\Delta_1$ = {10, 30, 80} ms and a b-value of 60 ms/μm$^2$, are G = {9.3, 5.3, 3.2} T/m; for the SDE time-dependence analysis, to achieve a b-value of 8 ms/μm$^2$ for a diffusion time of $\Delta_2$ = 1.1 ms the gradient strength needs to be G = 12.1 T/m; for the DDE sequences with $\Delta_3$ = 5 ms and $b_3$ = 16 ms/μm$^2$ the corresponding gradient is G = 4.9 T/m. Although these values are higher than what is practically feasible at present, similar SDE / DDE acquisitions can be achieved for example on high performance pre-clinical gradients which reach up to ~ 3 T/m, by increasing the gradient duration δ to 2-3 ms [44, 110]. For systems with lower gradient capabilities, the pulse durations would need to be further increased to reach high b-values, and the contrast provided by the sequences would need to be further investigated given the specific hardware constrains [64].

Future works will also benefit from the promising results obtained here, by, for example, using them to design real experiments targeting the non-invasive mapping of cell processes branch order in different areas of the brain known to be comprised of neural cells with very different morphologies: for instance the cerebral cortex, mostly comprised of pyramidal neurons, and the cerebellar molecular layer, mostly comprised of Purkinje cells' dendritic trees having branch order at least double that of pyramidal neurons.

## 5. Conclusion

This study uses advanced numerical simulations to systematically investigate for the first time the effects of dendritic branching on the dMRI and dMRS signal and shows the potential of DDE rather than SDE acquisitions to non-invasively map such microstructural features. In particular, the simulation results reported here can inform the design of dMRI/dMRS experiments focused on the quantification of branching order of cellular projections, a tissue feature of pivotal importance for characterizing a wide range of disorders [6], as well as normal and atypical development and aging [7]. Although a purely simulation study, our results are in good agreement with previously published dMRI and dMRS experimental evidence, supporting the fascinating perspective of non-invasively mapping the complex brain cell morphology in-vivo with double diffusion encoding measurements.

## Acknowledgments

The authors acknowledge support from the EPSRC grant no. EP/N018702/1 and the UKRI Future Leaders Fellowship grant no. MR/T020296/1 (to MP). AI's work also received support from "la

Caixa" Foundation (ID 100010434) and from the European Union's Horizon 2020 research and innovation programme under the Marie Skłodowska-Curie grant agreement No 847648 and No 101003390, fellowship code CF/BQ/PI20/11760029.

# Figures

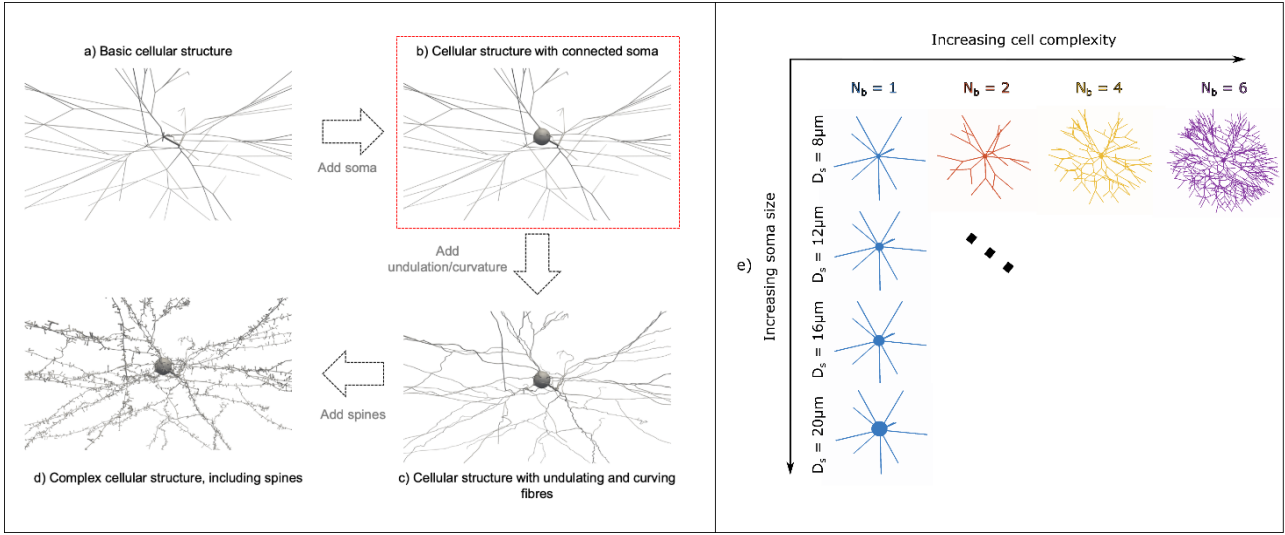

*Figure 1a)-d) Exemplar computational models of brain cell structures possible to design using the generative model introduced in [46]. The specific kind of cellular model used in this study is b) as highlighted by the red box. a) In this example, a basic cellular structure can be made using $N_{proj}$ = 10 interconnected cellular projections that can bifurcate in $N_b$ = 4 consecutive embranchments. Each branch has diameter $D_b$ = 0.75 μm and length $L_b$ = 125 μm. b) It is possible to make the cell model more complex, for example adding a cell body, namely soma, of given diameter, here $D_s$ = 20 μm, and/or c) adding branch undulations (direct over path ratio η = 0.95, see [46] for further details) and curvature (radius of curvature $R_c$ = 500 μm, see [46] for further details). d) Finally, for higher level of realism, secondary fine structural features, such as spines, can also be added choosing the density $ρ_{sp}$ = 2 spines/μm and the size of the spine head and neck, $h_{sp}$ = 0.5 μm and $n_{sp}$ = 1μm, respectively (see [46] for further details). e) Examples of synthetic cells with different branching orders, $N_b$ = {1,2,4,6} and different soma diameters $D_s$ = {8, 12, 16, 20} μm combinations used in this study. Cell morphological features other than soma size and branching order (as parametrized by $N_b$), such as undulations, curvature and spines, have been removed by design using the generative model in [46].*

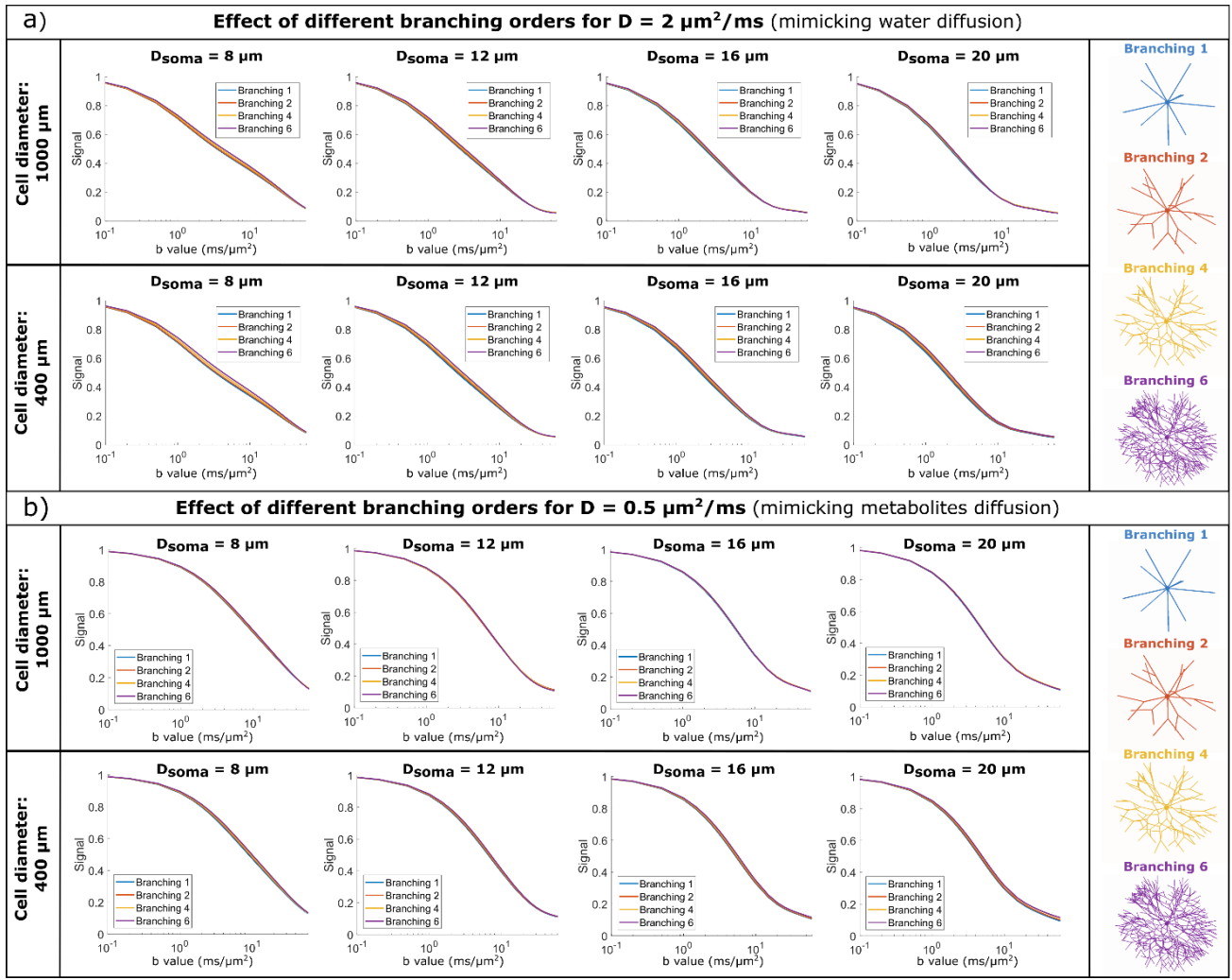

*Figure 2 B-value dependence of the diffusion signal for a) D = 2 μm²/ms (mimicking water diffusion) and b) D = 0.5 μm²/ms (mimicking metabolites diffusion). Within each panel, the top row simulates large cells and the bottom row small cells, with different soma diameters (increasing diameter from left to right) and various branching orders (lines of different colours). The data is simulated at Δ = 80 ms.*

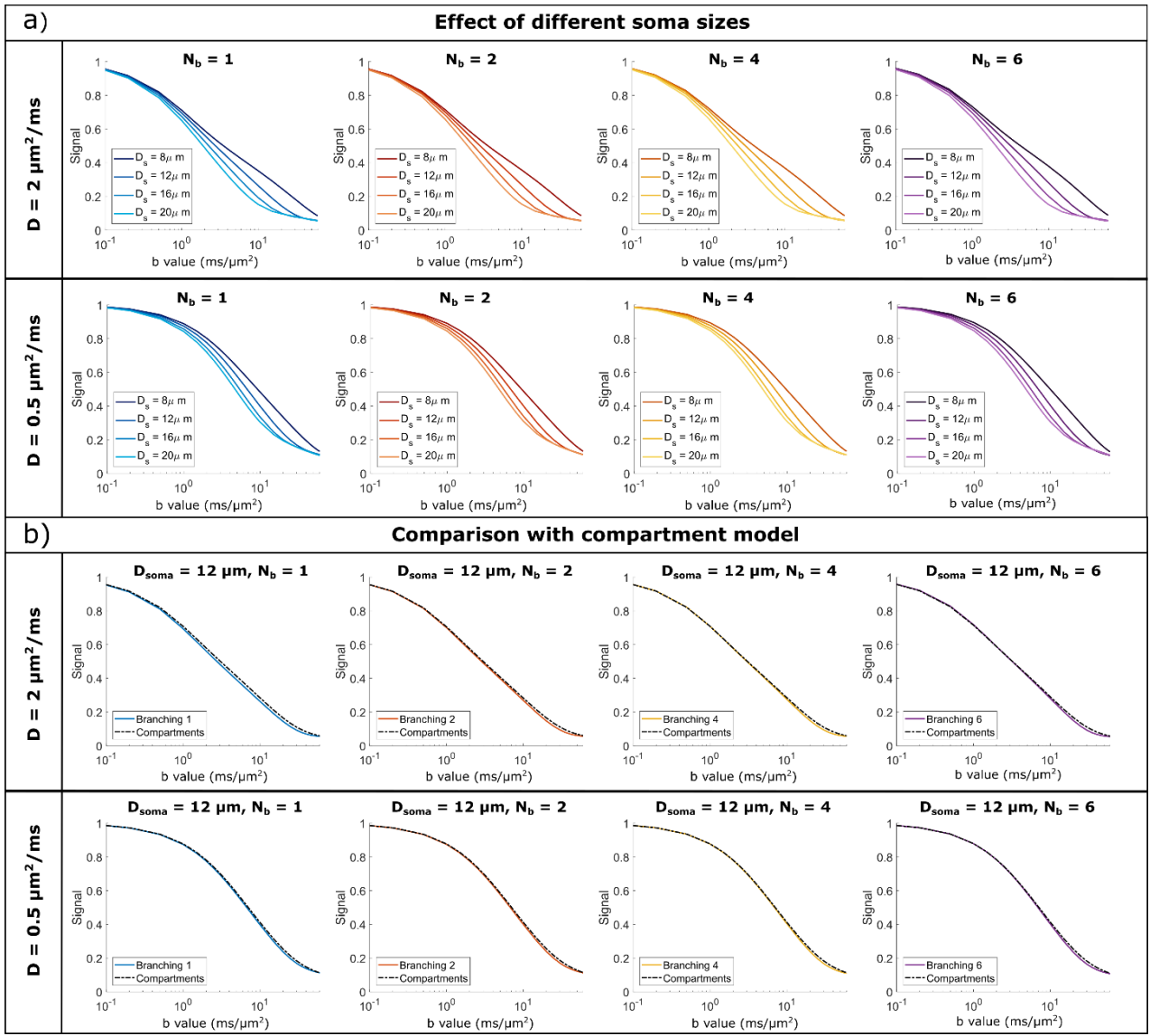

*Figure 3 a) Comparison of signal b-value dependence for cells with different soma sizes, domain size L = 400 μm, and increasing complexity (branching order increases left to right). b) Comparison between Monte Carlo simulations and the GPD approximation for a non-exchanging two-compartment model which includes diffusion inside a sphere and finite isotropically oriented cylinders with the same effective diameters as the effective $D_s$ and $D_b$ values indicated in Table 1. The signal is computed for cells with a domain size L = 400 μm and target $D_s$ = 8 μm. In both a) and b), the signal is computed for a diffusion time of 80 ms.*

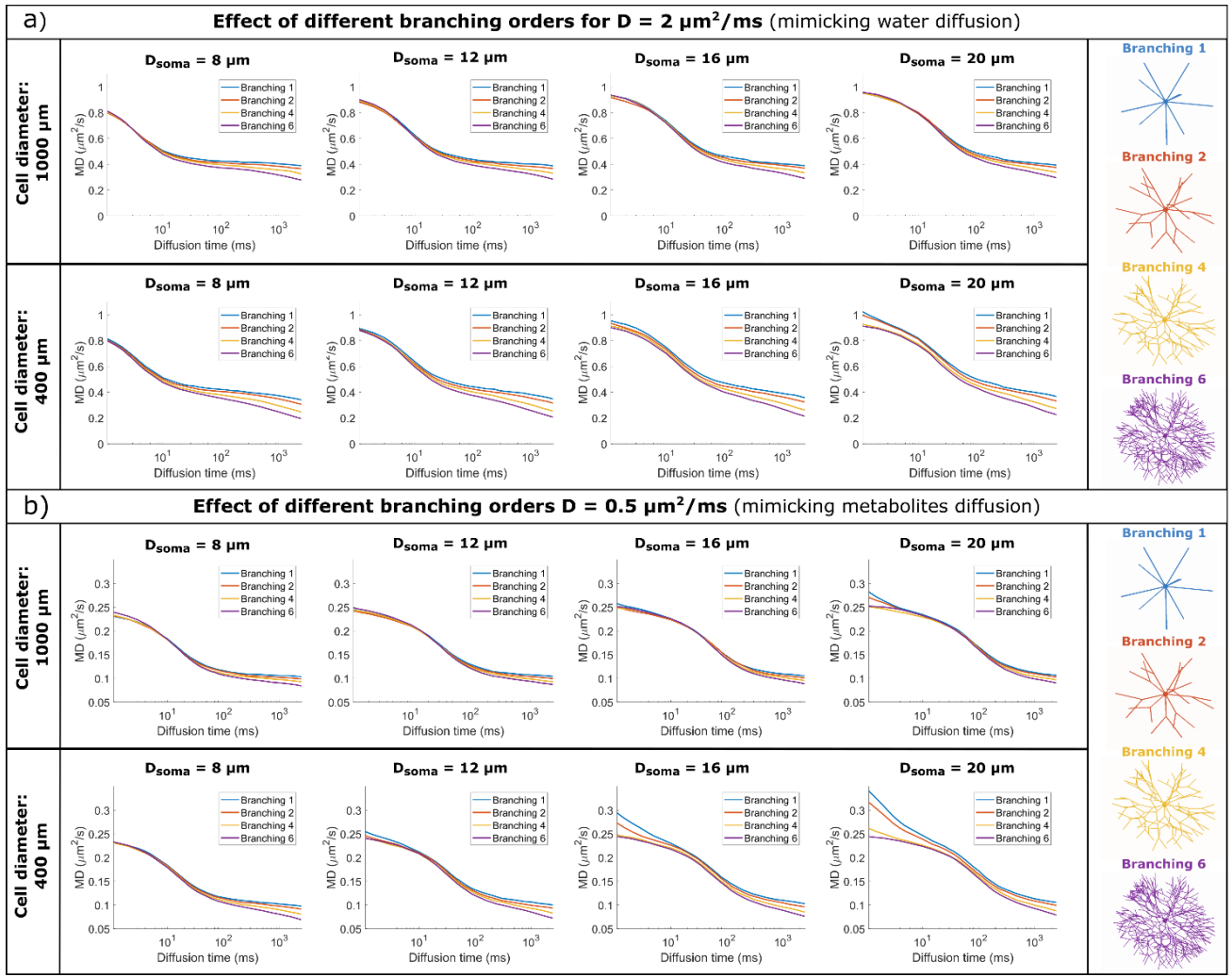

*Figure 4 MD time dependence for a) D = 2 μm²/ms (mimicking water diffusion) and b) D = 0.5 μm²/ms (mimicking metabolites diffusion). Within each panel, the top row simulates large cells and the bottom row small cells, with different soma diameters (increasing diameter from left to right) and various branching orders (lines of different colours). MD is computed by fitting the Kurtosis model to the data simulated with all 3 b-values.*

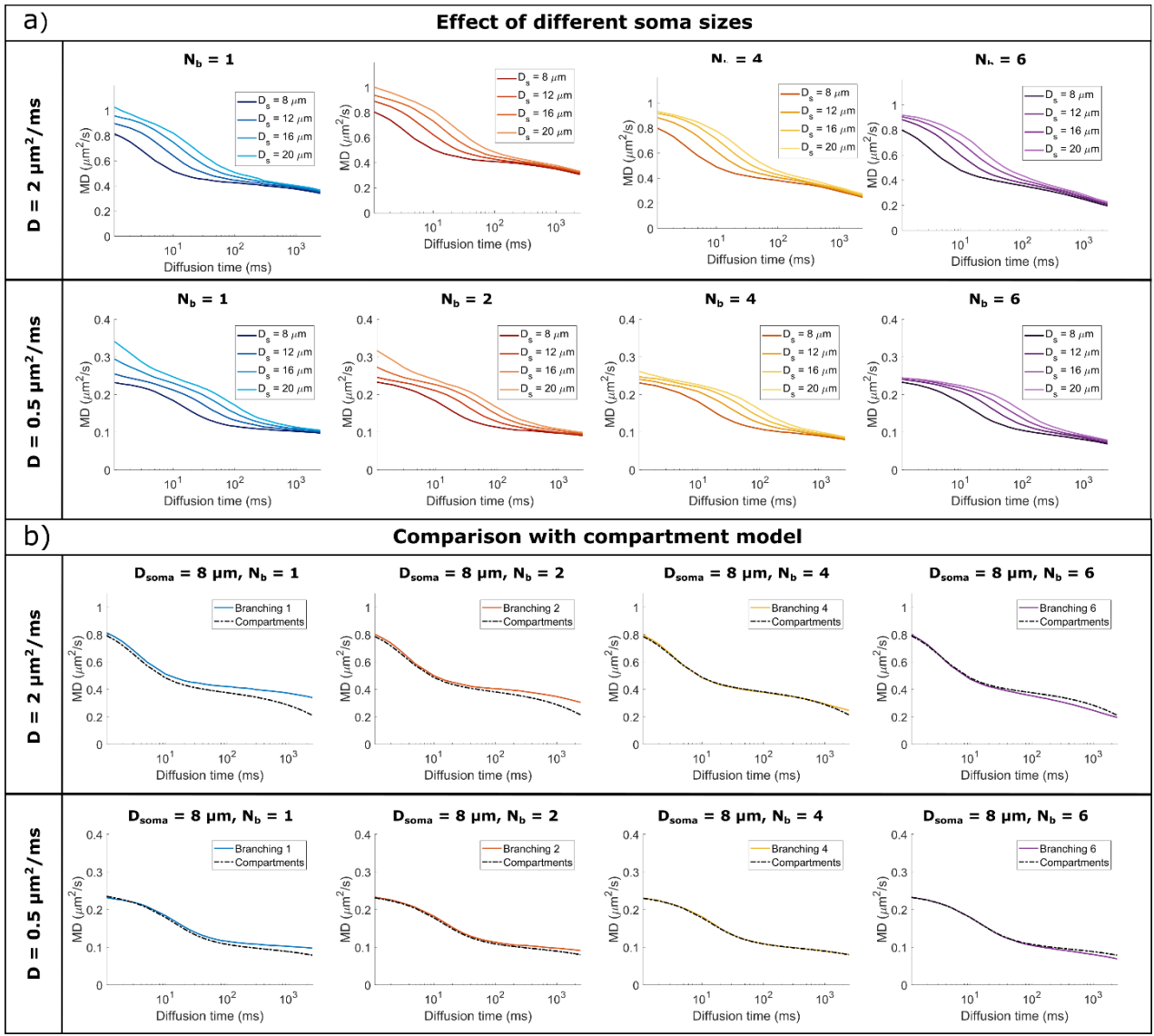

*Figure 5 a) Comparison of MD time dependence for cells with different soma sizes, domain size L = 400 μm, and increasing complexity (branching order increases left to right). b) Comparison between Monte Carlo simulations and the GPD approximation for a non-exchanging two-compartment model which includes diffusion inside a sphere and finite isotropically oriented cylinders with the same effective diameters as the effective $D_s$ and $D_b$ values indicated in Table 1. The signal is computed for cells with a domain size L = 400 μm and target $D_s$ = 8 μm. In both a) and b). MD is computed by fitting the Kurtosis model to the data simulated with all 3 b-values.*

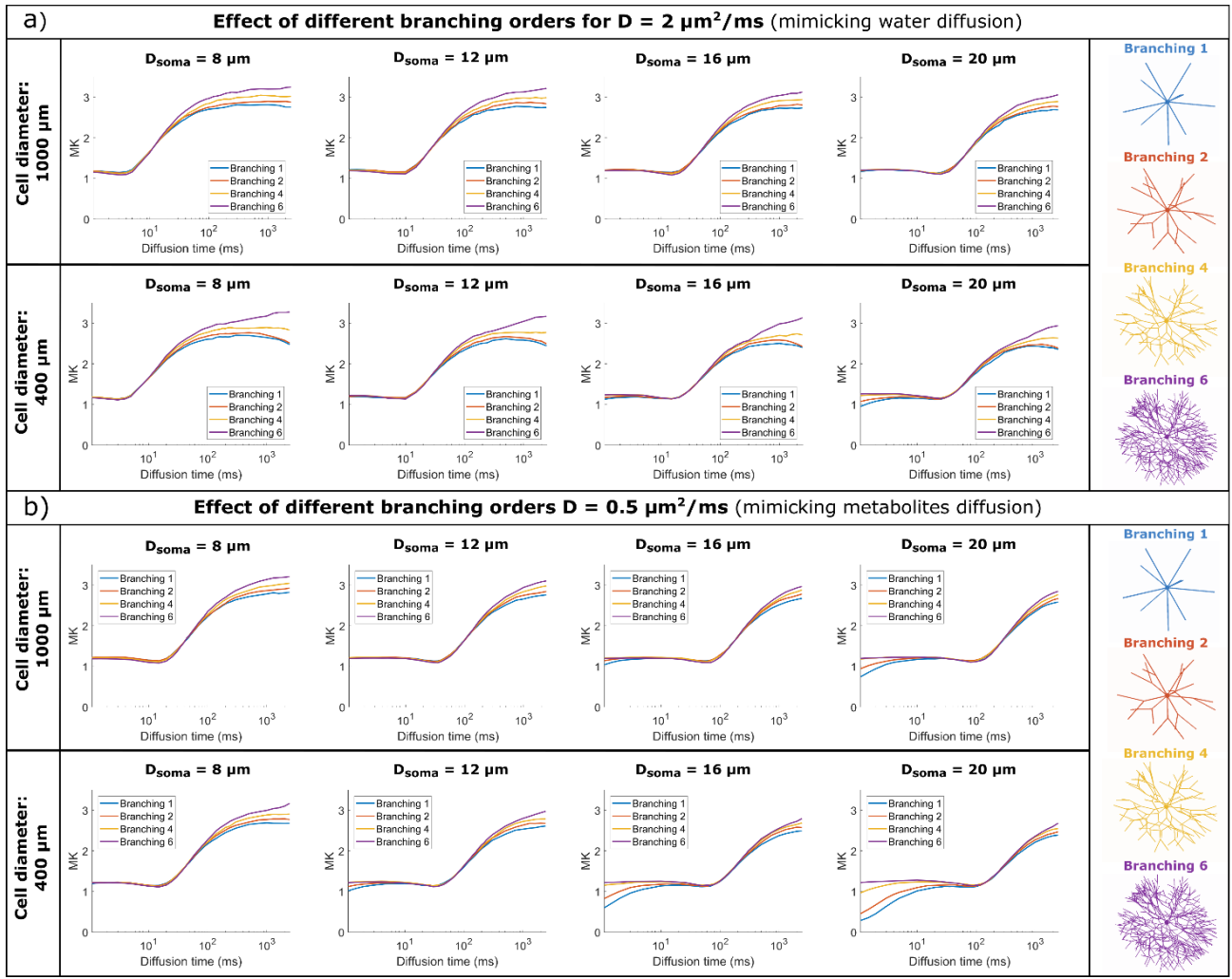

*Figure 6. MK time dependence for a) $D = 2$ $\mu m^2/ms$ (mimicking water diffusion) and b) $D = 0.5$ $\mu m^2/ms$ (mimicking metabolites diffusion). Within each panel, the top row simulates large cells and the bottom row small cells, with different soma diameters (increasing diameter from left to right) and various branching orders (lines of different colours). MK is computed by fitting the Kurtosis model to the data simulated with all 3 b-values.*

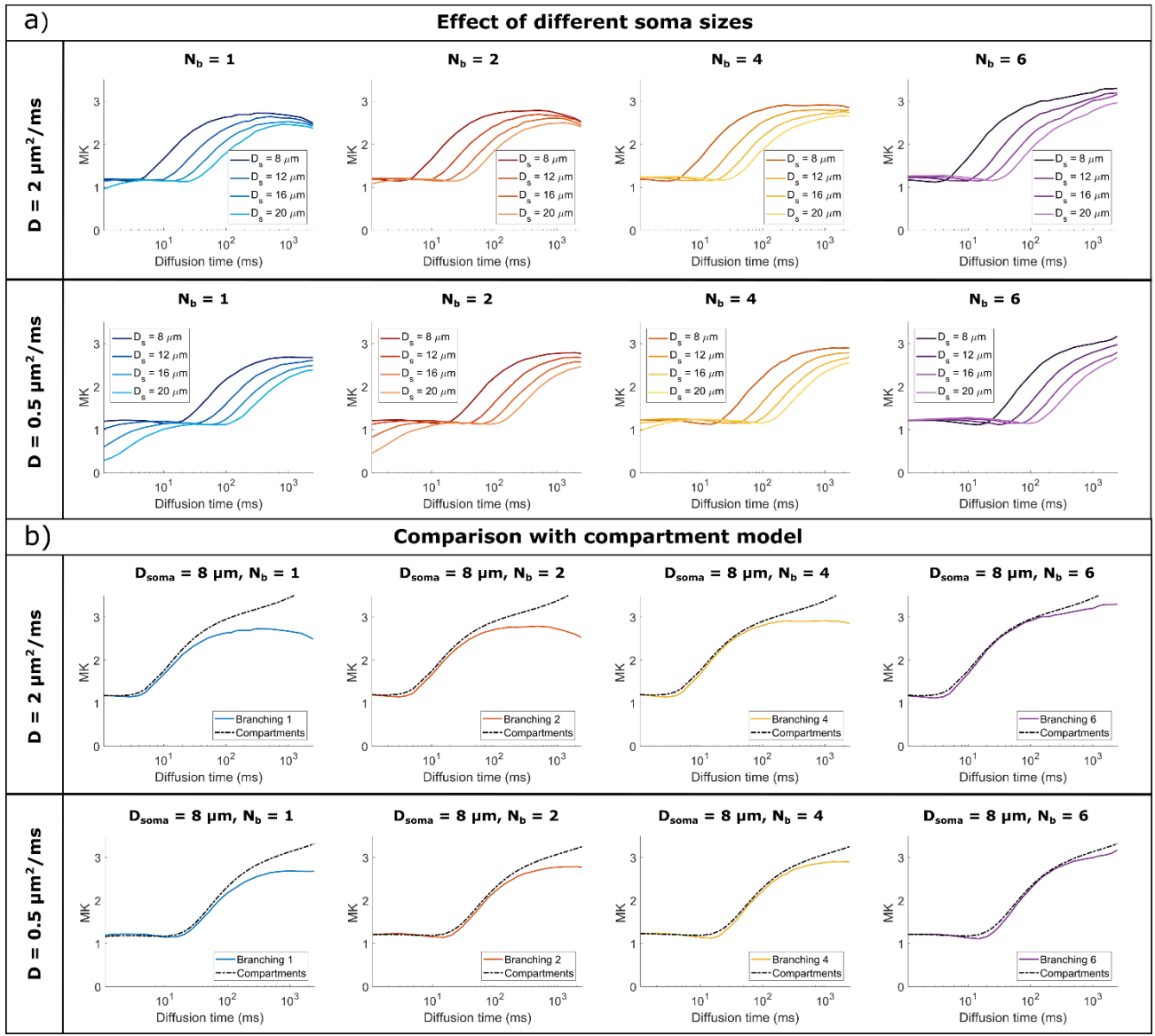

*Figure 7 a) Comparison of MK time dependence for cells with different soma sizes, domain size L = 400 μm, and increasing complexity (branching order increases left to right). b) Comparison between Monte Carlo simulations and the GPD approximation for a non-exchanging two-compartment model which includes diffusion inside a sphere and finite isotropically oriented cylinders with the same effective diameters as the effective $D_s$ and $D_b$ values indicated in Table S1. The signal is computed for cells with a domain size L = 400 μm and target $D_s$ = 8 μm. In both a) and b). MK is computed by fitting the Kurtosis model to the data simulated with all 3 b-values.*

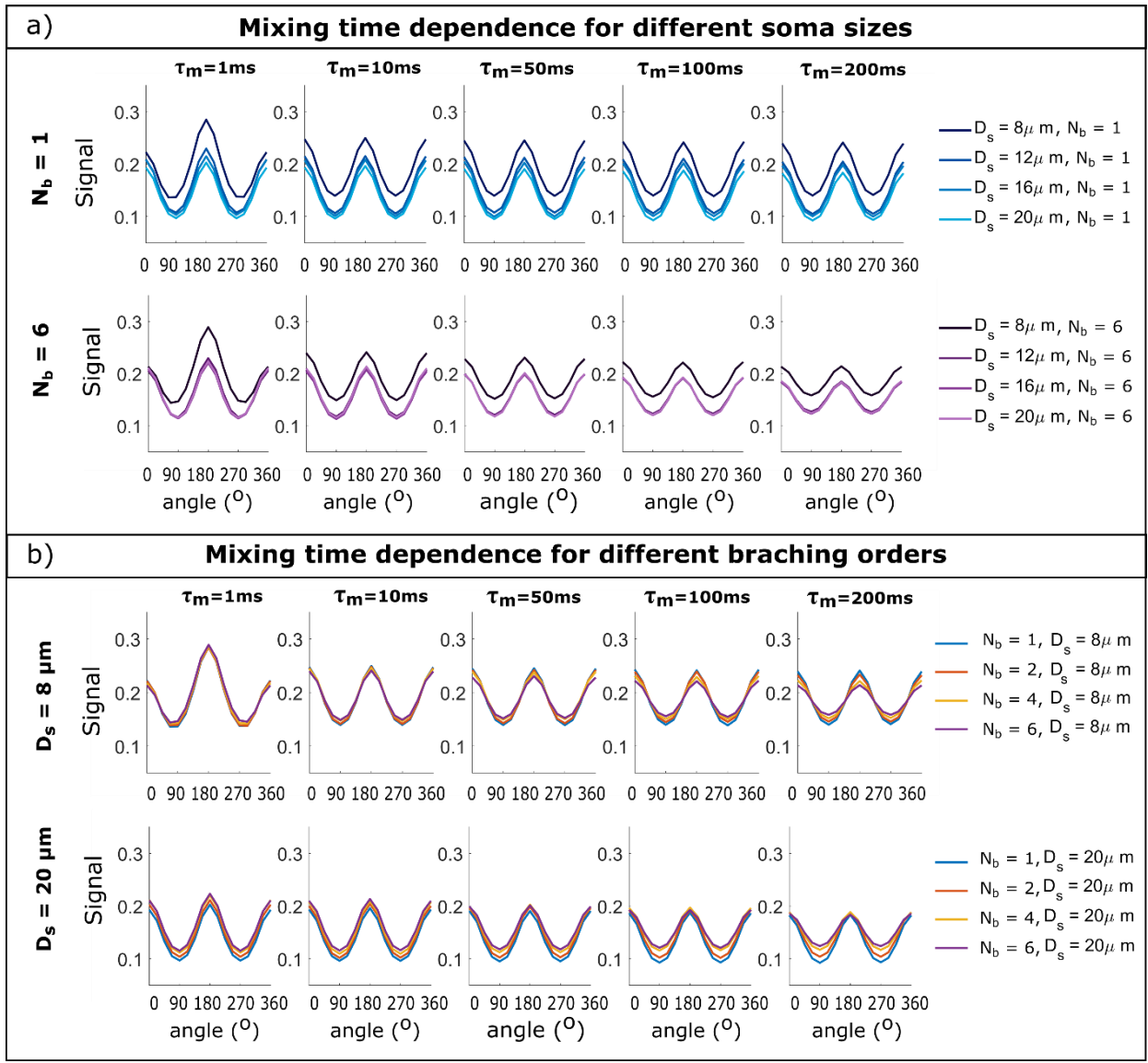

*Figure 8 Dependence of DDE angular modulation on the mixing time for a) cells with different soma sizes and b) cells with different branching orders. The data is simulated for cells with a domain size of 400 μm, b-value of b = 4 ms/μm² and D = 2 μm²/ms.*

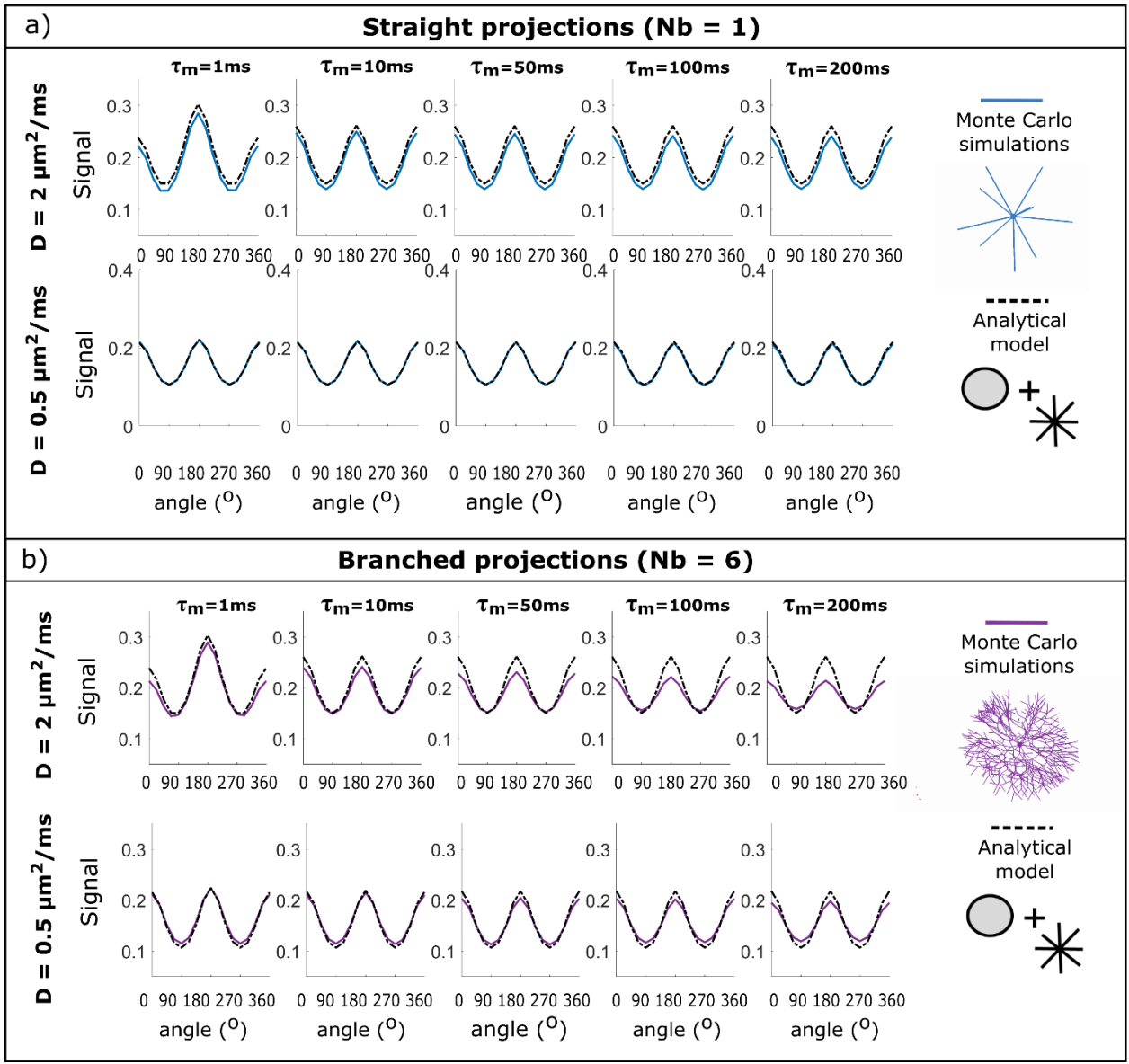

*Figure 9 Dependence of DDE angular modulation on the mixing time for cells with a soma diameter of 8 μm and a) $N_b = 1$ and b) $N_b = 6$. The data is simulated for cells with a domain size of 400 μm and a b-value of b = 4 ms/μm². The coloured solid lines represent the MC simulations, and the dotted lines represented the analytical two-compartment model.*

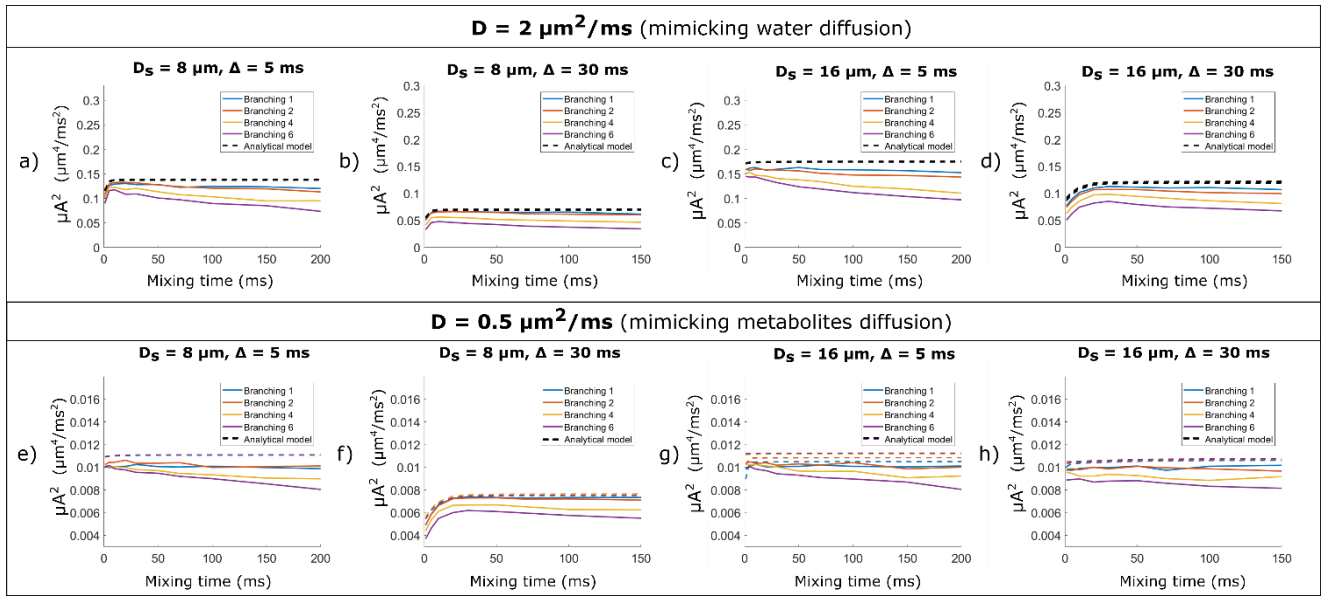

*Figure 10 Mixing time dependence of the apparent microscopic anisotropy for DDE sequences with $b = 4$ ms/μm$^2$ and different diffusion times and soma diameters: a),e) $\Delta = 5$ ms and $D_s = 8$ μm; b),f) $\Delta = 30$ ms and $D_s = 8$ μm; c),g) $\Delta = 5$ ms and $D_s = 16$ μm; d),h) $\Delta = 30$ ms and $D_s = 16$ μm. The data is simulated for diffusivities a)-d) $D = 2$ μm$^2$/ms and e)-h) $D = 0.5$ μm$^2$/ms. The data is simulated for cells with an overall diameter of 400 μm. The coloured solid lines represent the MC simulations, and the dotted lines represented the analytical two-compartment model.*

# Supplementary Information

| | $N_b$ | $L_b$ (μm) | Target $D_s$ (μm) | Effective $D_s$ (μm) | $v_s$ | Target $D_b$ (μm) | Effective $D_b$ (μm) | $p_{ex}$ (%) |
|---|---|---|---|---|---|---|---|---|
| a) Overall cell diameter = 1000 μm | 1 | 500 | 8 | 7.73 | 0.32 | 0.45 | 0.36 | 0.54 |
| | 1 | 500 | 12 | 11.60 | 0.32 | 0.83 | 0.66 | 0.81 |
| | 1 | 500 | 16 | 15.47 | 0.32 | 1.28 | 1.03 | 1.11 |
| | 1 | 500 | 20 | 19.36 | 0.32 | 1.78 | 1.44 | 1.38 |
| | 2 | 250 | 8 | 7.73 | 0.32 | 0.37 | 0.29 | 0.35 |
| | 2 | 250 | 12 | 11.59 | 0.31 | 0.67 | 0.55 | 0.56 |
| | 2 | 250 | 16 | 15.47 | 0.31 | 1.04 | 0.84 | 0.74 |
| | 2 | 250 | 20 | 19.34 | 0.31 | 1.46 | 1.19 | 0.95 |
| | 4 | 125 | 8 | 7.72 | 0.32 | 0.22 | 0.18 | 0.14 |
| | 4 | 125 | 12 | 11.58 | 0.32 | 0.41 | 0.34 | 0.22 |
| | 4 | 125 | 16 | 15.44 | 0.32 | 0.64 | 0.53 | 0.29 |
| | 4 | 125 | 20 | 19.32 | 0.31 | 0.90 | 0.75 | 0.38 |
| | 6 | 83 | 8 | 7.72 | 0.34 | 0.12 | 0.10 | 0.04 |
| | 6 | 83 | 12 | 11.58 | 0.34 | 0.23 | 0.19 | 0.07 |
| | 6 | 83 | 16 | 15.44 | 0.33 | 0.36 | 0.30 | 0.09 |
| | 6 | 83 | 20 | 19.30 | 0.33 | 0.51 | 0.43 | 0.12 |
| | $N_b$ | $L_b$ (μm) | Target $D_s$ (μm) | Effective $D_s$ (μm) | $v_s$ | Target $D_b$ (μm) | Effective $D_b$ (μm) | $p_{ex}$ (%) |
| b) Overall cell diameter = 400 μm | 1 | 200 | 8 | 7.74 | 0.32 | 0.71 | 0.57 | 1.36 |
| | 1 | 200 | 12 | 11.62 | 0.32 | 1.31 | 1.06 | 2.08 |
| | 1 | 200 | 16 | 15.52 | 0.31 | 2.02 | 1.65 | 2.83 |
| | 1 | 200 | 20 | 19.42 | 0.31 | 2.82 | 2.32 | 3.57 |
| | 2 | 100 | 8 | 7.74 | 0.31 | 0.58 | 0.47 | 0.92 |
| | 2 | 100 | 12 | 11.61 | 0.31 | 1.07 | 0.88 | 1.44 |
| | 2 | 100 | 16 | 15.50 | 0.3 | 1.65 | 1.38 | 1.98 |
| | 2 | 100 | 20 | 19.39 | 0.3 | 2.30 | 1.95 | 2.53 |
| | 4 | 50 | 8 | 7.73 | 0.31 | 0.36 | 0.30 | 0.38 |
| | 4 | 50 | 12 | 11.59 | 0.3 | 0.66 | 0.56 | 0.58 |
| | 4 | 50 | 16 | 15.47 | 0.3 | 1.01 | 0.88 | 0.81 |
| | 4 | 50 | 20 | 19.34 | 0.29 | 1.42 | 1.26 | 1.06 |
| | 6 | 33 | 8 | 7.72 | 0.32 | 0.20 | 0.17 | 0.12 |
| | 6 | 33 | 121 | 11.58 | 0.31 | 0.37 | 0.33 | 0.2 |
| | 6 | 33 | 16 | 15.44 | 0.3 | 0.57 | 0.52 | 0.28 |
| | 6 | 33 | 20 | 19.32 | 0.29 | 0.80 | 0.74 | 0.37 |

*Table S2 Parameters of the computational models of cellular meshes used in MC simulations.*

S1. Additional simulation results

In the first part, we present additional simulation results for different cellular configurations and/or sequence parameters compared to the data shown in the main text which further support the results and discussion of this work.

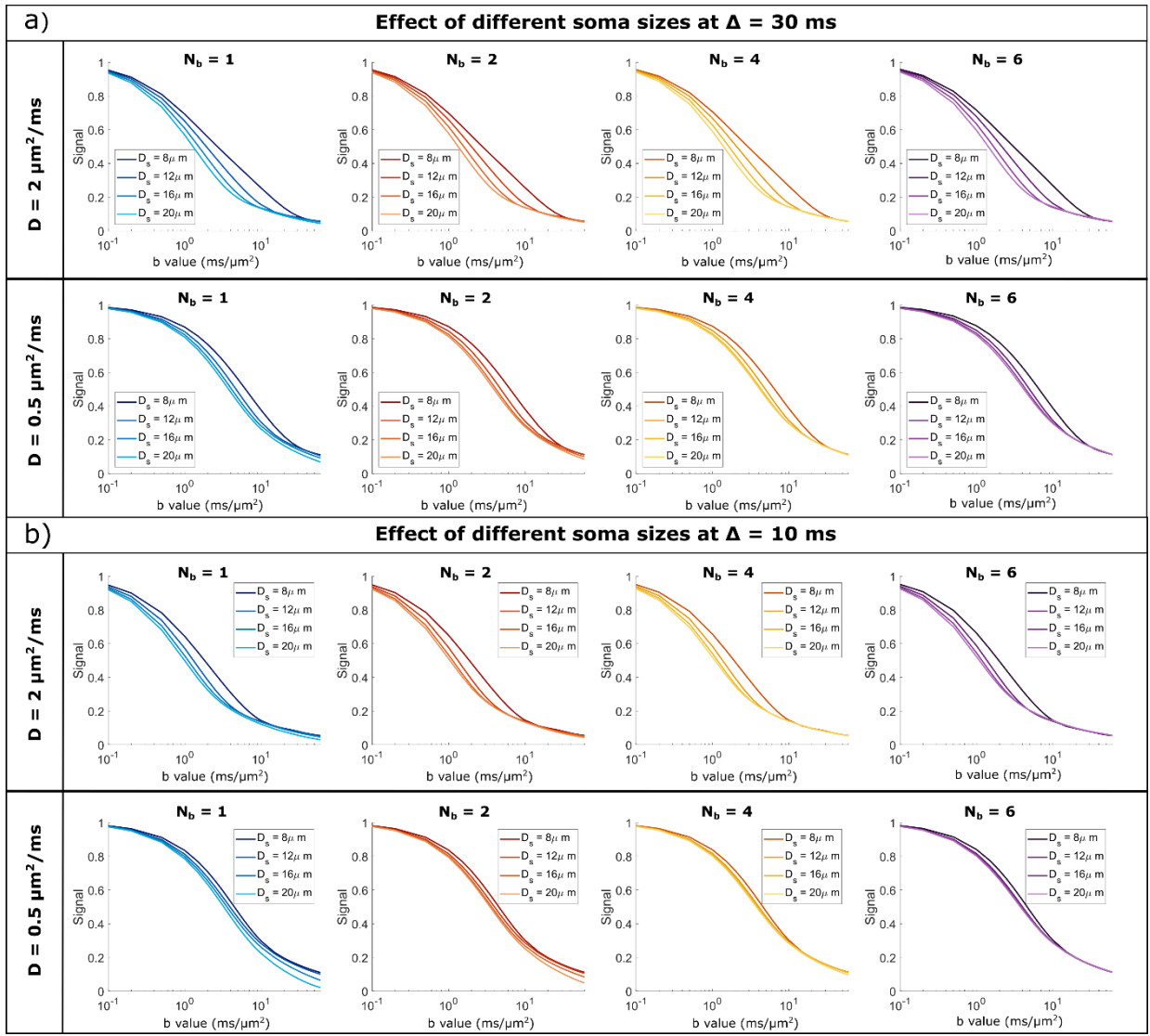

*Figure S1 b-value dependence SDE signal for cells with different soma sizes and branching orders for a diffusion time of a) Δ = 30 ms and b) Δ = 10 ms.*

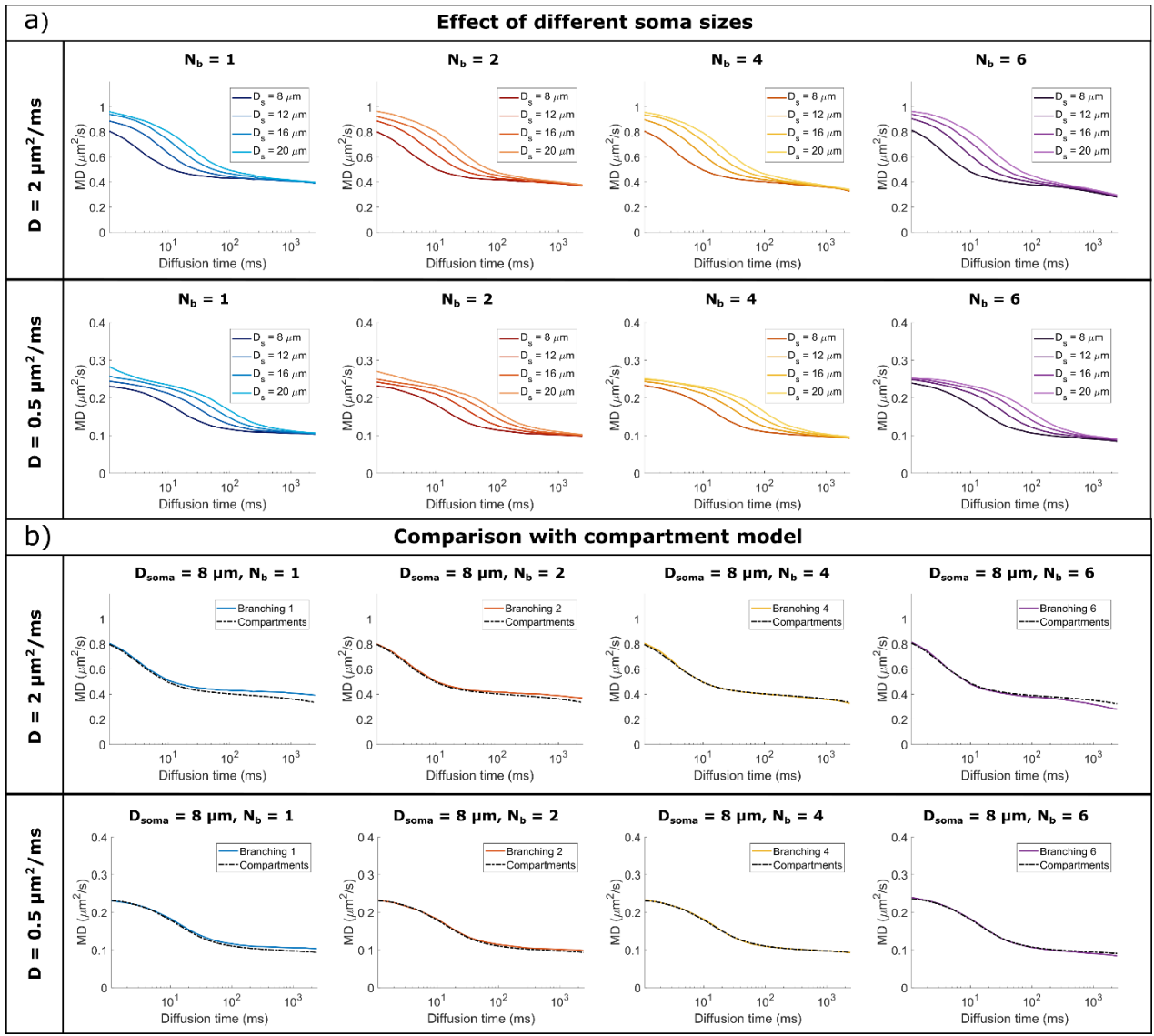

*Figure S2 a) Effect of different soma sizes and branching orders on the MD time dependence for cells with a domain L = 1000 µm. b) Comparison of MD time dependence between simulated data and a two-compartment model for cells with different branching orders, $D_s$ = 8 µm and L = 1000 µm.*

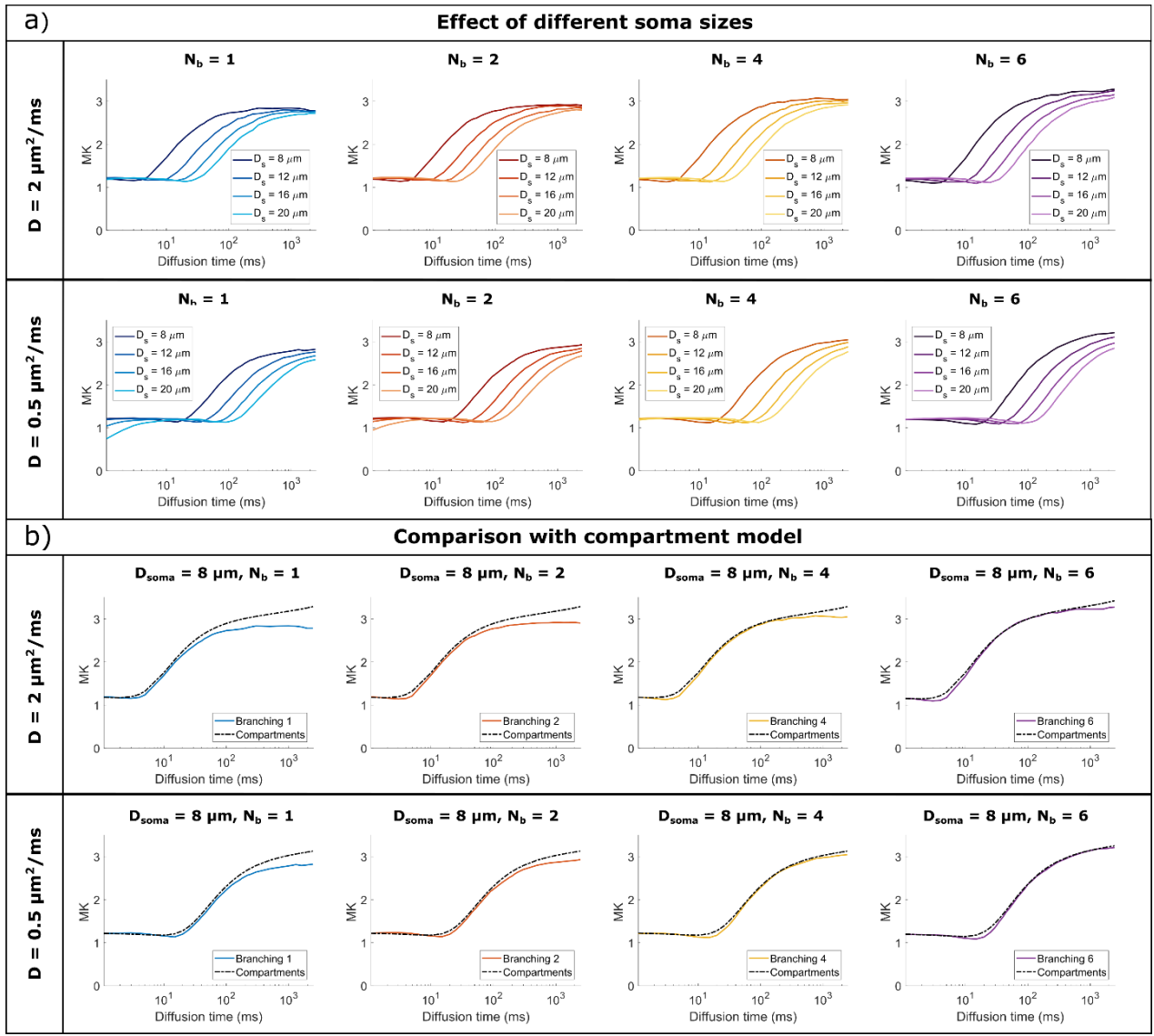

*Figure S3 a) Effect of different soma sizes and branching orders on the MK time dependence for cells with a domain L = 1000 μm. b) Comparison of MK time dependence between simulated data and a two-compartment model for cells with different branching orders, $D_s$ = 8 μm and L = 1000 μm.*

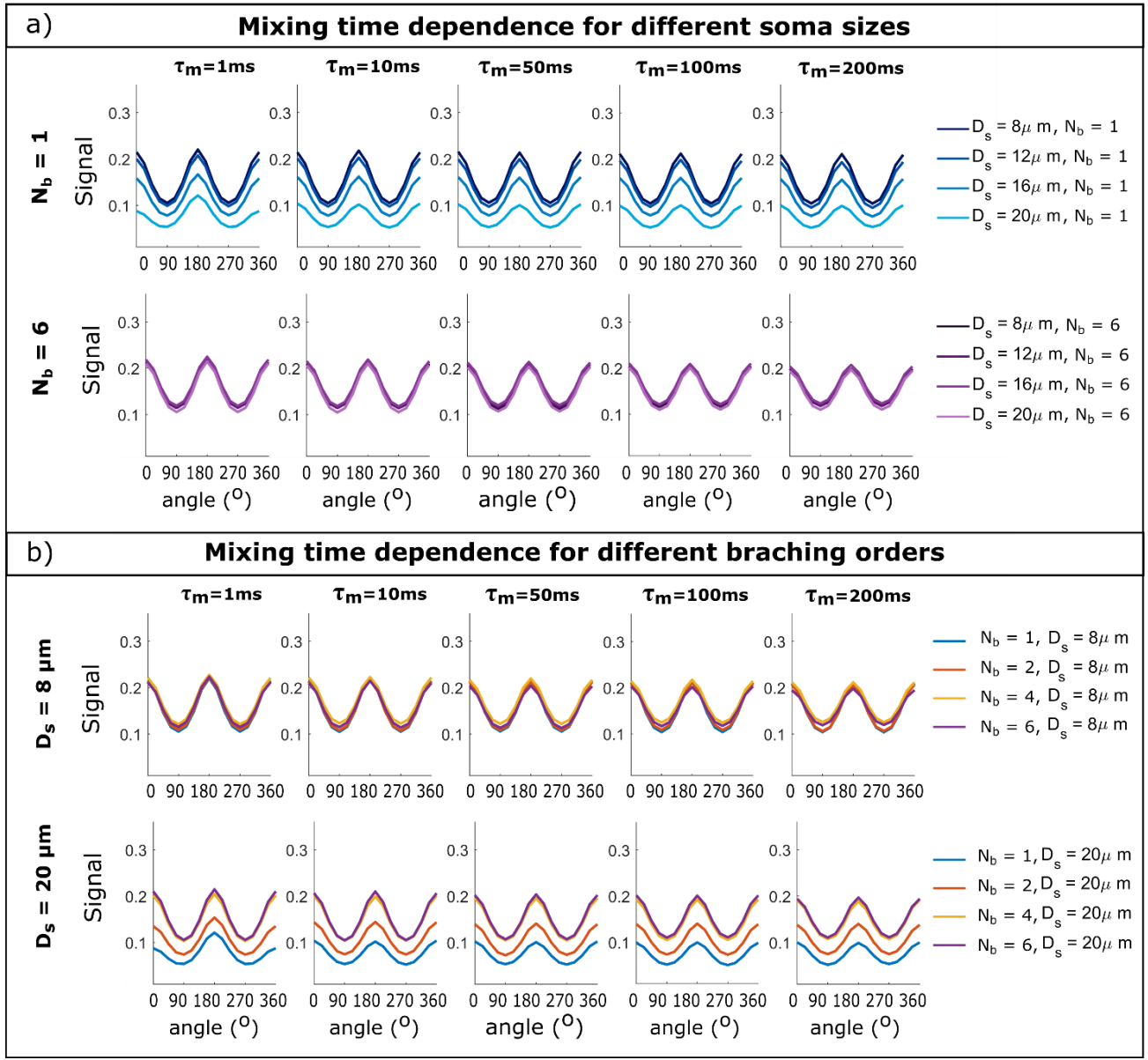

*Figure S4. Dependence of DDE angular modulation on the mixing time for a) cells with different soma sizes and b) cells with different branching orders. The data is simulated for cells with a domain size of 400 μm, b-value of b = 16 ms/μm² and D = 0.5 μm²/ms.*

S2. Analysis of signal differences in the presence of noise

In this part, we present the impact of soma size and cell complexity on the SDE and DDE signal from the perspective of detectability at different noise levels. To this end, after the signal was averaged over the 10 cellular configurations, $N_{noise}$ = 1000 instances of Gaussian noise with standard deviation σ = 0.05 (i.e. corresponding to an SNR of 20 in the b0 data) was added to each diffusion measurement. Then, the signal was averaged over directions, as described for each experiment. Following the directional average, the estimated metrics were computed.

For SDE sequences we investigate the signal differences as a function of b-value, as well as differences in MD and MK as a function of diffusion time, similar to the analysis presented in section 2.3, after noise was added to the data.

Figures S5a) and S6a) present the signal difference between configurations with the smallest diameter $D_s = 8$ μm and those with larger diameters, as a function of b-value for a diffusion time of 80 and 10 ms, respectively. Figures S5b) and S6b) present the signal difference between the MC simulations and the theoretical compartment model for cells of various soma size and branching orders, for a diffusion time of 80 and 10 ms, respectively.

Figures S7a) and S8a) present the MD and MK differences between configurations with the smallest diameter Ds = 8 μm and those with larger diameters, as a function of diffusion time, while Figures S7b) and S8b) present the MD and MK differences between the MC simulations and the theoretical compartment model, for the same parameters as the data presented in Figures 6 and 7.

For DDE sequences, we investigate the mixing time dependence of the amplitude of the signal modulation between measurements with parallel and orthogonal gradients. To calculate the amplitude modulation, we first compute the mean signal for measurements with parallel and anti-parallel gradients (i.e. measurements with $\varphi = 0$ and $\pi$ in section 2.4.1) and then we subtract the mean signal for measurements with orthogonal gradients (i.e. measurements with $\varphi = \pi/2$ and $3\pi/2$ in section 2.4.1). To see whether changes are detectable, we analyse the difference in amplitude modulation between measurements with increasing mixing times and $\tau_m = 1$ ms, in the presence of noise. Thus, after the signal was averaged over the 10 cellular configurations, we add $N_{noise} = 1000$ instances of Gaussian noise with standard deviation $\sigma = 0.05$ (i.e. corresponding to an SNR of 20 in the b0 data) to each diffusion measurement. Then, for each relative angle, the signal was averaged over the 8 different planes and the 5 in-plane rotations. After averaging, the amplitude modulation was computed as described above.

Figure S9a) presents the difference in amplitude modulation between measurements with increasing mixing time and $\tau_m = 1$ ms for cells with different soma diameters and branching orders. Figure S9b) illustrates the difference in amplitude modulation between the MC simulations and the theoretical compartment model for cells with different soma diameters and branching orders.

To further investigate the effect of noise on the mixing time dependence of the estimated apparent microscopic anisotropy, we add $N_{noise} = 1000$ instances of Gaussian noise with standard deviation $\sigma = 0.05$ (i.e. corresponding to an SNR of 20 in the b0 data) to each diffusion measurement from the 5-design protocol employed in section 2.5.2. Figure S10 illustrates the difference in apparent μA between the MC simulations and the theoretical compartment model as a function of mixing time for cells with different soma diameters and branching orders, for DDE sequences with $\Delta = 5$ ms, and $\Delta = 30$ ms.

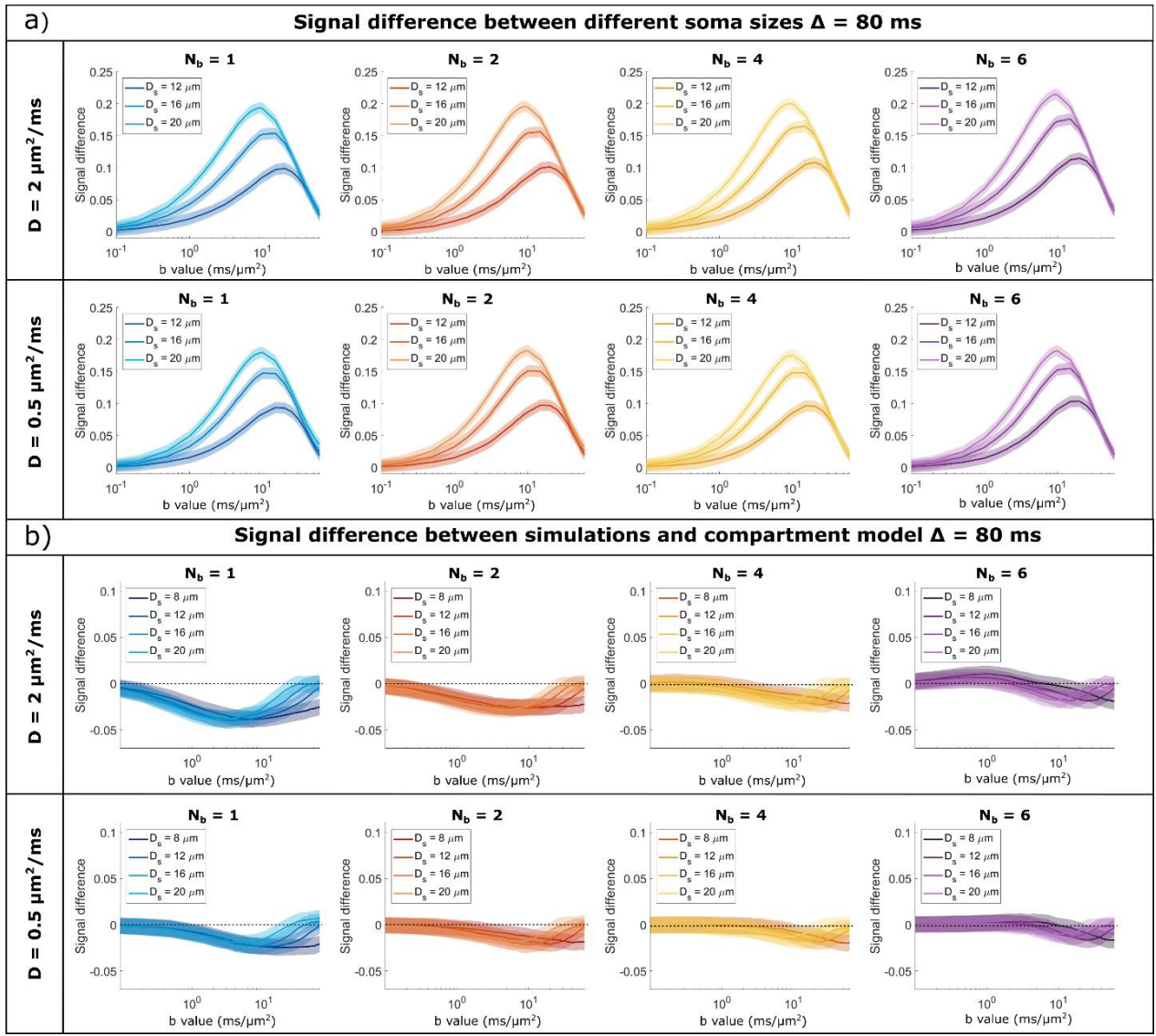

*Figure S5 a) Signal difference between cells with larger soma diameters (12 – 20 μm) and cells with Ds = 8 μm as a function of b-value, for cells with Nb = {1, 2, 4, 6} and L = 400 μm. b) Signal difference between the MC simulations and the compartment model for cells with different soma diameters and branching orders. The shaded area represents the standard deviation over 1000 noisy datapoints. When the shaded areas do not overlap, the differences are detectable, meaning that they are statistically significant with a p<0.01. The data is simulated at Δ = 80 ms.*

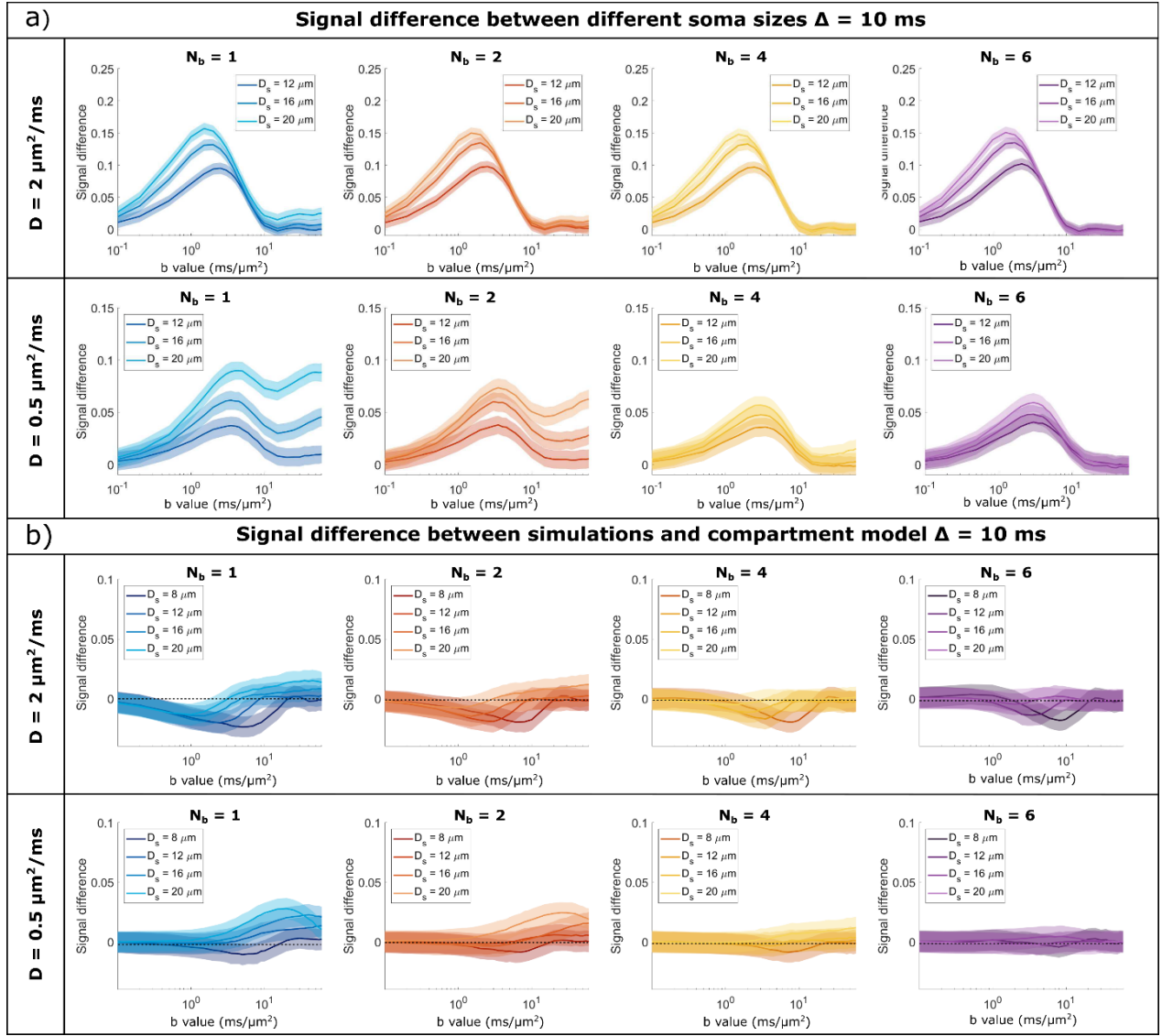

*Figure S6 a) Signal difference between cells with the smallest diameter Ds = 8 μm and those with larger diameters (12 – 20 μm) as a function of b-value, for cells with Nb = {1, 2, 4, 6} and L = 400 μm. b) Signal difference between the MC simulations and the compartment model for cells with different soma diameters and branching orders. The shaded area represents the standard deviation over 1000 noisy datapoints. When the shaded areas do not overlap, the differences are detectable, meaning that they are statistically significant with a p<0.01. The data is simulated at Δ = 10 ms.*

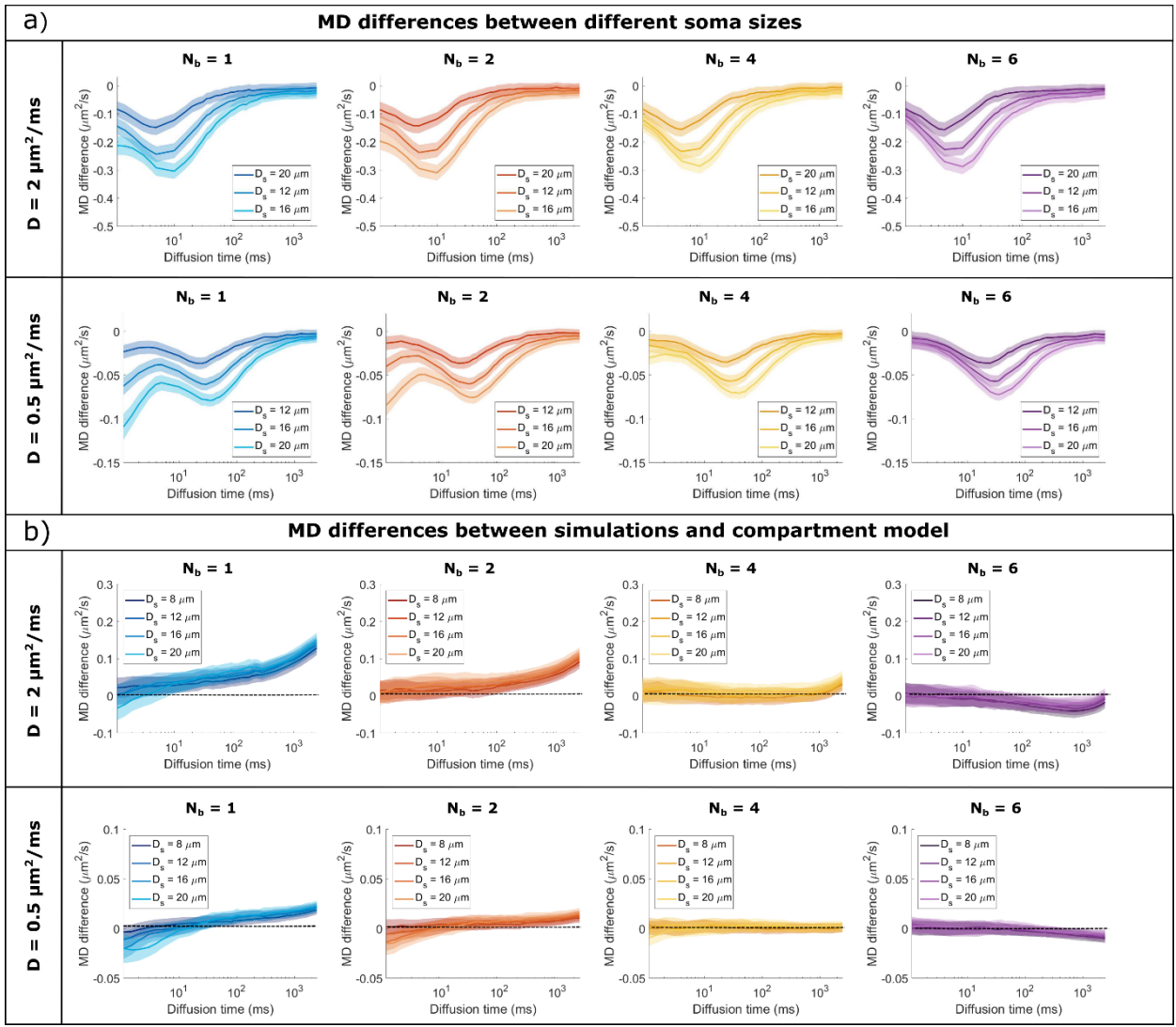

*Figure S7 a) MD differences between cells with the smallest diameter Ds = 8 μm and those with larger diameters (12 – 20 μm) as a function of diffusion time, for cells with Nb = {1, 2, 4, 6}[1] and L = 400 μm. b) MD differences between the MC simulations and the compartment model for cells with different soma diameters and branching orders. The shaded area represents the standard deviation over 1000 noisy datapoints. When the shaded areas do not overlap, the differences are detectable, meaning that they are statistically significant with a p<0.01.*

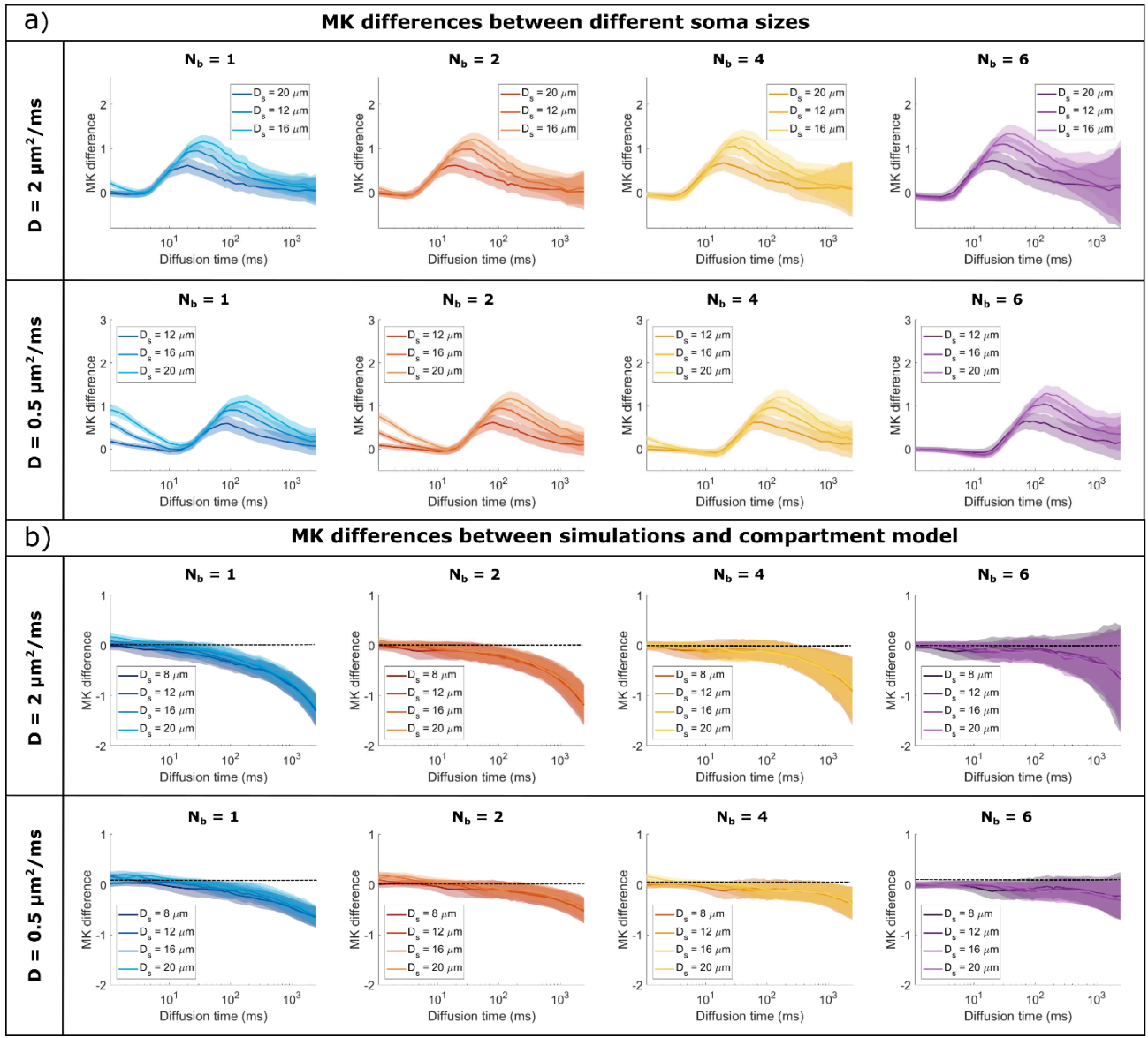

*Figure S8 a) MK differences between cells with the smallest diameter Ds = 8 µm and those with larger diameters (12 – 20 µm) as a function of diffusion time, for cells with Nb = {1, 2, 4, 6}[1] and L = 400 µm. b) MK differences between the MC simulations and the compartment model for cells with different soma diameters and branching orders. The shaded area represents the standard deviation over 1000 noisy datapoints. When the shaded areas do not overlap, the differences are detectable, meaning that they are statistically significant with a p<0.01.*

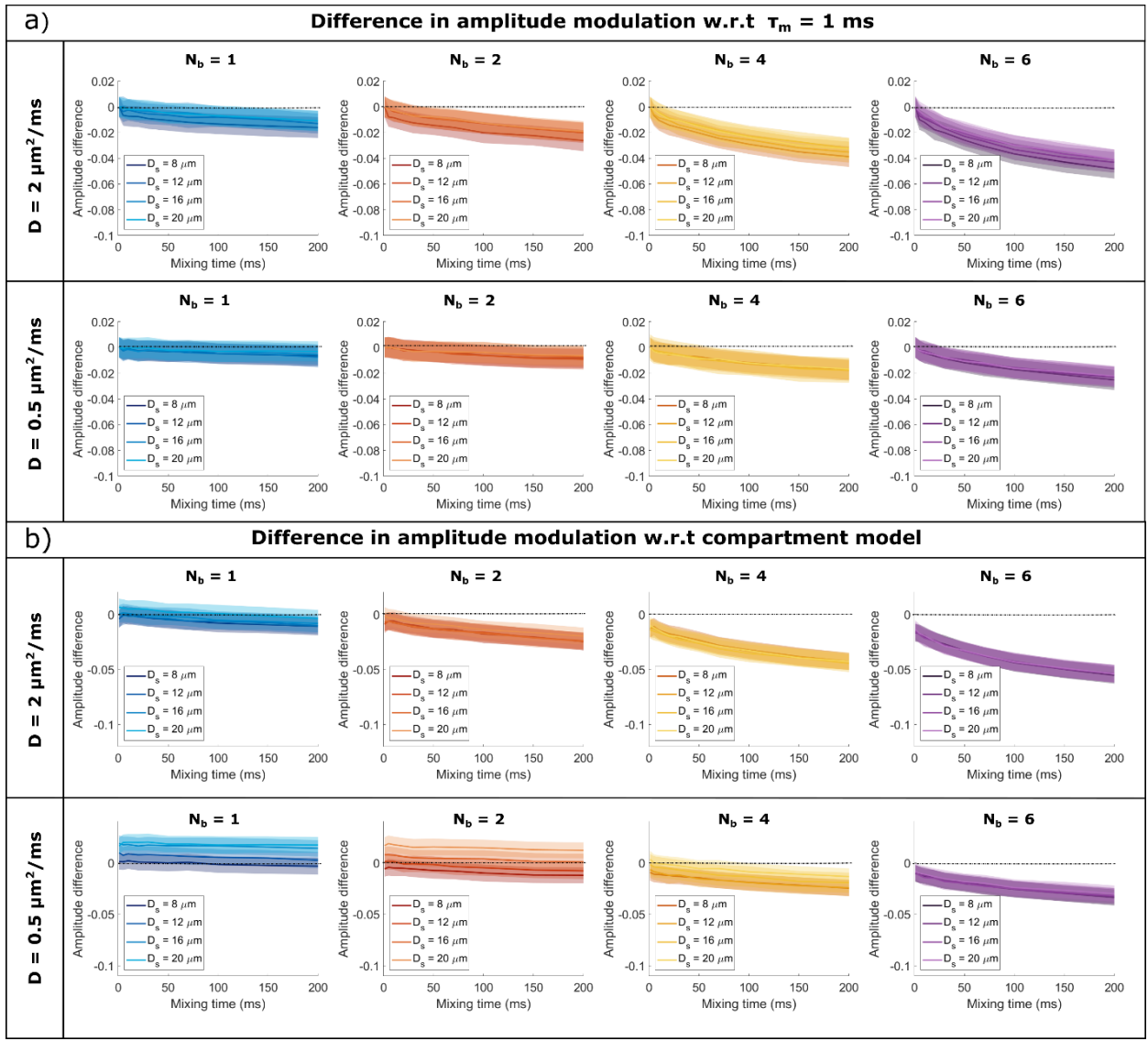

*Figure S9a) Difference in amplitude modulation between measurements with increasing mixing time and $\tau_m = 1$ ms for cells with different soma diameters and branching orders. b) difference in amplitude modulation between the MC simulations and the theoretical compartment model for cells with different soma diameters and branching orders. The simulations have been performed with the same parameters as detailed in Section 2.4.1 ($\Delta = 5$ ms, $b = 4$ ms/µm² for $D = 2$ µm²/ms and $b = 16$ ms/µm² for $D = 0.5$ µm²/ms). The shaded area represents the standard deviation over 1000 noisy datapoints. When the shaded areas do not overlap, the differences are detectable, meaning that they are statistically significant with a $p<0.01$.*

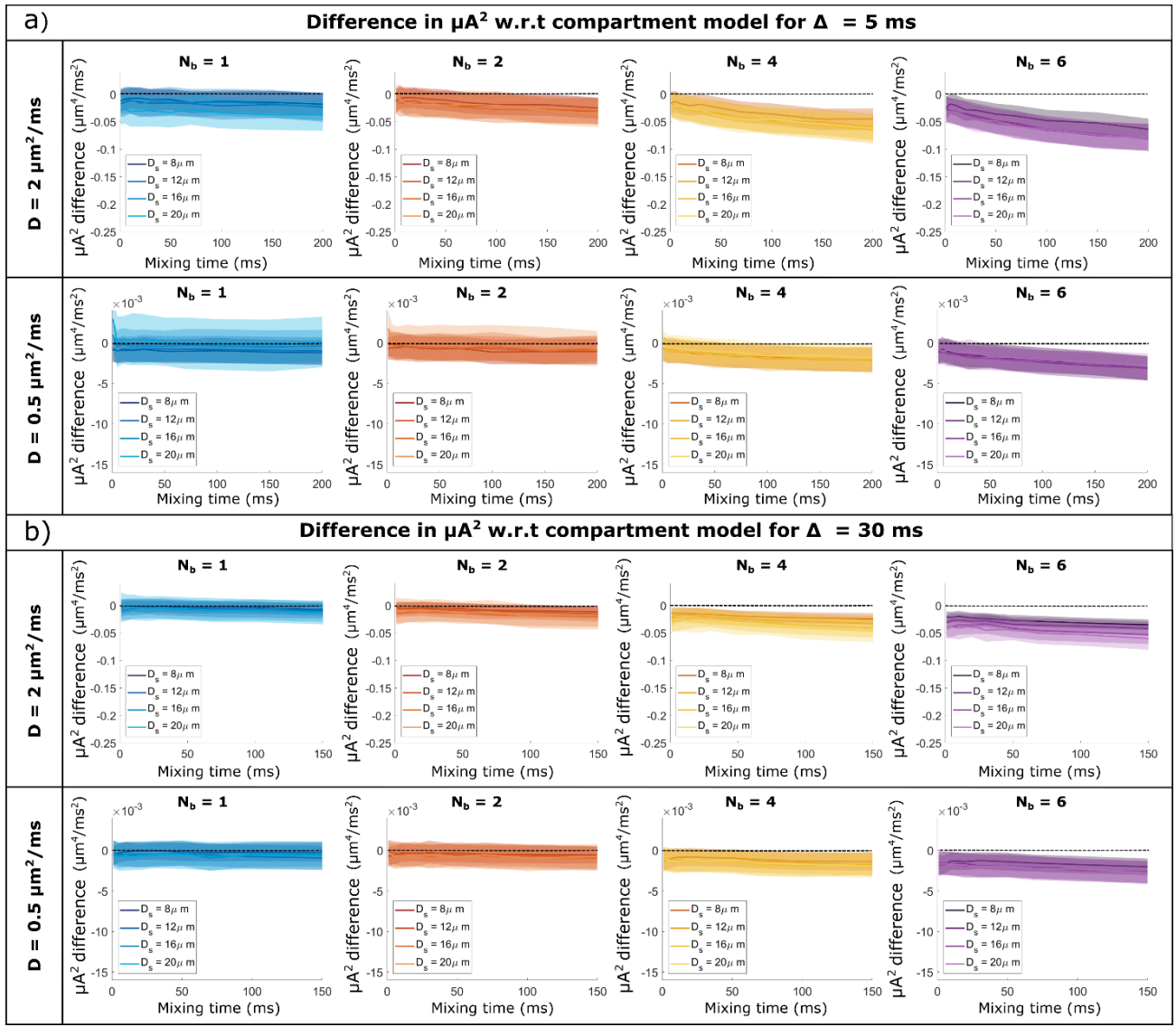

*Figure S10a) Difference in µA² between the MC simulations and the theoretical compartment model for cells with different soma diameters and branching orders for DDE with Δ = 5 ms. The simulations have been performed with the same parameters as detailed in Section 2.5.2 (b = 4 ms/µm² for D = 2 µm²/ms and b = 16 ms/µm² for D = 0.5 µm²/ms) . S10b) Same as a), just for Δ = 30 ms. The shaded areas represent the standard deviation of the estimated metrics over 1000 instances of noise. When the shaded areas do not overlap, the differences are detectable, meaning that they are statistically significant with a p<0.01.*

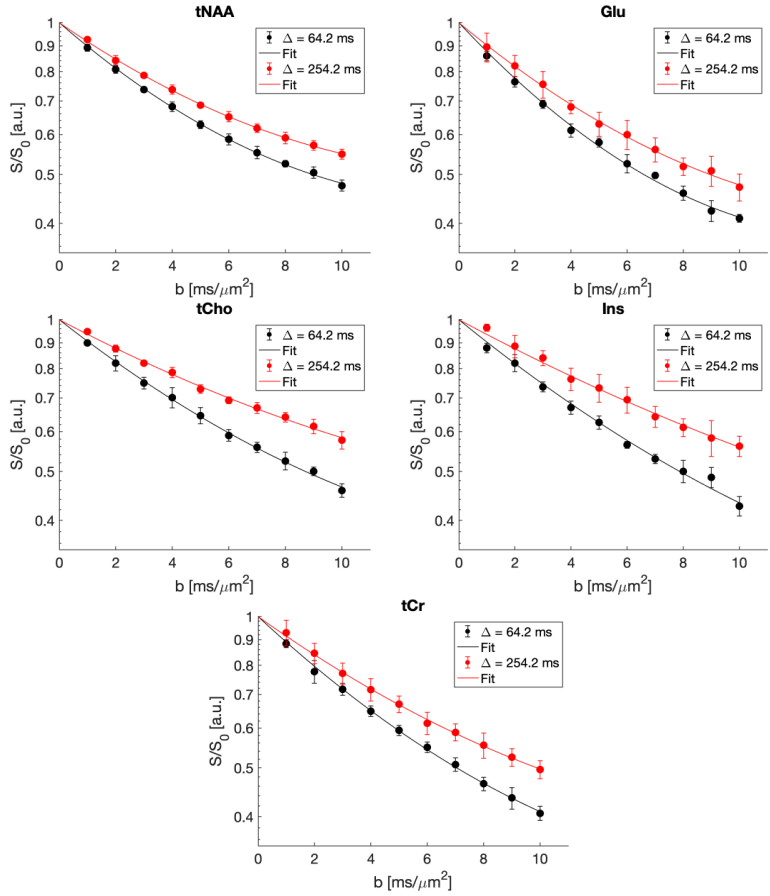

*Figure S11. dMRS data from [5] analysed using the cumulant expansion of the diffusion weighted signal up to the second order. The datapoints are average signal values over four mice and the error bars are the corresponding standard deviations. The solid lines are the fit of the second-order cumulant expansion representation (i.e., Eq.[4] in the main text) to the data. The estimated MD and MK values for each diffusion gradient separation Δ are reported in the table, with the 95% confidence interval reported in square brackets. More information about the data acquisition and processing can be found in [5].*

### S3. Quantification of exchange effects from the AXR model

In this analysis we aim to quantify the impact of exchange between two compartments on the DDE signal using the simple apparent exchange rate (AXR) model. Following equations 6 and 7 in [2] we can calculate the DDE signal as a function of b-value and mixing time as:

$$S(b, \tau_m) = S_0(\tau_m) \exp(-b\, ADC'(\tau_m)),$$

where $S_0(\tau_m)$ is the signal at b = 0 and accounts for the effects of longitudinal relaxation during the mixing time and the apparent diffusion coefficient $ADC'(\tau_m)$, in the limit $b \to 0$, is determined by the apparent diffusivity and exchange rate of the two compartments:

$$ADC'(\tau_m) = ADC(1 - \sigma \exp(-\tau_m AXR)),$$

where $ADC = f_1^e D_1 + (1 - f_1^e) D_2$ is the equilibrium apparent diffusion coefficient of the system, σ is the filter efficiency and AXR is the apparent exchange rate. From previous literature the estimated parameter values (ADC, σ and AXR) depend on the tissue [1, 3, 4], but also on the b-value of the first gradient pair, with [4] reporting a drastic decrease in AXR between values estimated with a filter of b = 250 s/mm$^2$ and b = 900 s/mm$^2$. Reported ADC values were between ~ 0.6 to 0.8 for WM and ~

0.8 and 1 in GM [1, 3] [4]; σ values estimated between ~0.2 and 0.3 in both WM and GM; AXR values were between 0.4 and 0.8 in GM and between 0.6 and 0.9 in WM, for a filter b-value of 900 s/mm$^2$.

To estimate the impact on the DDE sequences simulated in this work, we calculate the DDE signal difference between measurements with $\tau_m$ = 1 ms and $\tau_m$ = 200 ms for two combinations of parameter values measured in GM: σ = 0.23, ADC = 0.75 mm$^2$/s and AXR = 0.72 s$^{-1}$ from [4] and σ = 0.23, ADC = 0.95 mm$^2$/s and AXR = 0.4 s$^{-1}$ from [1]. Assuming $S_0(\tau_m)$ is constant and the same parameters apply for a filter with b = 2000 s/mm$^2$, we obtain normalized signal differences due to exchange effects of 0.009 and 0.004, respectively. These differences might become even smaller given the decrease in AXR which has been reported when the filter was increased from 250 to 900 s/mm$^2$. Even with the current values the signal differences due to exchange as modelled by AXR are >5 times smaller than the differences quantified in Figure S7 due to increasing the branching order.